\documentclass[preprint,12pt]{elsarticle}

\usepackage{lineno}
\usepackage{graphicx}
\usepackage{gensymb}
\usepackage{color}
\usepackage{subfig}
\usepackage{textcomp}
\usepackage{amssymb}
\usepackage{mathtools}
\usepackage{tcolorbox}
\usepackage{siunitx}
\usepackage{hepnames}
\usepackage{multirow}
\usepackage{hyperref}
\usepackage[capitalise, noabbrev]{cleveref}

\usepackage{pgfplots}
\usepackage{tikz}
\pgfplotsset{compat=1.5.1}
\usetikzlibrary{fadings}
\usepackage{pgffor}

\newcommand{\SpacePoints}{space points}
\newcommand{\SpacePoint}{space point}
\newcommand{\SpacePointCapital}{Space point}
\newcommand{\SectorMap}{Sector Map}
\newcommand{\SectorMaps}{Sector Maps}

\newcommand{\BelleII}{Belle~II\xspace}
\newcommand{\genfit}{GENFIT2\xspace}
\newcommand{\FourMomentum}{four\nobreakdash-momentum\xspace}
\newcommand{\percentageNumber}[1]{{#1}\%}

\begin{document}

\begin{frontmatter}
\title{Track Finding at \BelleII}
\author{\BelleII Tracking Group
}

\address[a:pisa]{
  Scuola Normale Superiore and INFN Sezione di Pisa, Largo B. Pontecorvo 3, Pisa, 56127, Italy
}

\address[a:prag] {
  Charles Univ. Prague, Institute of Particle and Nuclear Physics, Ke Karlovu 3, Praha, 12116, Czechia 
}

\address[a:kit]{
  Karlsruhe Institute of Technology (KIT), Institute of Experimental Particle Physics (ETP), Wolfgang-Gaede-Str. 1, Karlsruhe, D-76131, Germany
}

\address[a:pisa2]{
  Dipartimento di Fisica, Universit\`{a} di Pisa and INFN Sezione di Pisa, Largo B. Pontecorvo 3, Pisa, 56127 Italy
}

\address[a:desy]{
Deutsches Elektronen-Synchrotron (DESY), Notkestrasse 85,
   Hamburg, D-22607, Germany
}

\address[a:lal]{
  Laboratoire de L'accelerateur Lineaire (LAL) Orsay, Batiment 200 BP34 91898 ORSAY Cedex, France
}

\address[a:rome]{
  Dipartimento di Matematica e Fisica, Universit\`{a} di Roma Tre and INFN Sezione di Roma Tre, Via della vasca navale 84, Roma, 00146, Italy
}

\address[a:fudan]{
  Fudan University, Key Laboratory of Nuclear Physics and Ion-beam Application (MOE)
  and Institute of Modern Physics, Shanghai, 200443, China
}

\address[a:stras]{
  Universit\'e de Strasbourg, CNRS, IPHC, UMR 7178, Strasbourg, 67037, France
}

\address[a:bonn]{
  Rheinische Friedrich-Wilhelms-Univ. Bonn, Physikalisches Institut, Nussallee 12, Bonn, D-53115, Germany
}

\address[a:ihepA]{
  Institut of High Energy Physics of the Austrian Academy of Sciences, Nikolsdorfer Gasse 18,
Wien, 1050, Austria 
}

\address[a:goett]{
  University of G\"ottingen, Faculty of Physics, Friedrich-Hund-Platz 1, G\"ottingen, D-37077, Germany
}

\address[a:lmu] {
  Ludwig Maximilians Univ. M\"unchen (LMU), Faculty of Physics, Geschwister-Scholl-Platz 1, Munich, D-80539, Germany
}

\address[a:ihepC]{
  Institute of High Energy Physics (IHEP), 19B Yuquan Road, Beijing, 100049, China
}

\address[a:kra] {
  Institute of Nuclear Physics PAN, Division of Particle Physics and Astrophysics, Department of Leptonic Interactions, ul. Radzikowskiego 152,
  Krakow, 31342, Poland
}

\address[a:nago]{
  Nagoya Univeristy, Graduate School of Science, Furou-cho Chikusa-ku, Aichi, Nagoya-shi, 464-8602, Japan
}

\address[a:iiser] {
  Indian Institute of Science Education and Research (IISER) Mohali, Knowledge city, Sector 81, SAS Nagar, Manauli, Punjab, 140306, India
}

\address[a:vpi] {
  Virginia Polytechnic Institute and State Univ., 315 A Robeson Hall, Blacksburg, VA 24061-0435, USA
}

\address[a:mainz]{
  Johannes Gutenberg Univ. of Mainz, Johann-Joachim-Becher-Weg 45, Mainz, D-55128, Germany
}

\address[a:lpt]{
  LP-Research, Inc., 
  Nishiazabu 1, Minato City,  Tokyo, 106-0031, Japan
}

\address[a:tori] {
  INFN and Univ. Torino, Via P. Giuria 1 I-10125, Torino, Italy
}

\address[a:tokyoU]{
  Department of Physics, Graduate School of Science, The University of Tokyo, 7-3-1 Hongo, Bunkyo-ku, Tokyo 113-0033, Japan
}

\address[a:melb] {
  University of Melbourne, School of Physics, Melbourne, Victoria 3010, Australia
}

%-----------------------------------------------
\author[a:pisa]{Valerio~Bertacchi}
\author[a:prag]{Tadeas~Bilka}
\author[a:kit]{Nils~Braun}
\author[a:pisa2]{Giulia~Casarosa}
\author[a:pisa2]{Luigi~Corona}
\author[a:desy]{Sam~Cunliffe}
\author[a:desy]{Filippo~Dattola}
\author[a:lal]{Gaetano~De~Marino}
\author[a:desy]{Michael~De~Nuccio}
\author[a:rome]{Giacomo~De~Pietro}
\author[a:fudan]{Thanh~Van~Dong}
\author[a:stras]{Giulio~Dujany}
\author[a:kit]{Patrick~Ecker}
\author[a:bonn]{Michael~Eliachevitch}
\author[a:stras]{Tristan~Fillinger}
\author[a:desy]{Oliver~Frost}
\author[a:ihepA]{Rudolf~Fr\"uhwirth}
\author[a:goett]{Uwe~Gebauer}
\author[a:desy]{Sasha~Glazov}
\author[a:lmu]{Nicolas~Gosling}
\author[a:desy,a:ihepC]{Aiqiang~Guo}
\author[a:kit]{Thomas~Hauth}
\author[a:kit]{Martin~Heck}
\author[a:kra]{Mateusz~Kaleta}
\author[a:prag]{Jakub~Kandra}
\author[a:desy]{Claus~Kleinwort}
\author[a:lmu]{Thomas~Kuhr}
\author[a:desy]{Simon~Kurz}
\author[a:prag]{Peter~Kvasnicka}
\author[a:ihepA]{Jakob~Lettenbichler}
\author[a:lmu]{Thomas~Lueck}
\author[a:rome]{Alberto~Martini}
\author[a:kit]{Felix~Metzner}
\author[a:nago]{Dmitrii~Neverov}
\author[a:desy]{Carsten~Niebuhr}
\author[a:pisa2]{Eugenio~Paoloni}
\author[a:iiser]{Sourav~Patra}
\author[a:vpi]{Leo~Piilonen}
\author[a:desy]{Cyrille~Praz}
\author[a:kit]{Markus~Tobias~Prim}
\author[a:kit]{Christian~Pulvermacher}
\author[a:kit]{Sebastian~Racs}
\author[a:desy]{Navid~Rad}
\author[a:desy]{Petar~Rados}
\author[a:lmu]{Martin~Ritter}
\author[a:pisa2]{Giuliana~Rizzo}
\author[a:desy]{Armine~Rostomyan}
\author[a:mainz]{Bianca~Scavino}
\author[a:lpt]{Tobias~Schl\"uter}
\author[a:goett]{Benjamin~Schwenker}
\author[a:tori]{Stefano~Spataro}
\author[a:mainz]{Bj\"orn~Spruck}
\author[a:desy]{Henrikas~Svidras}
\author[a:desy]{Francesco~Tenchini}
\author[a:tokyoU]{Yuma~Uematsu}
\author[a:melb]{James~Webb}
\author[a:bonn]{Christian~Wessel}
\author[a:pisa2]{Laura~Zani}

% !TeX encoding = UTF-8
% !TeX spellcheck = en_US
% !TeX root = ./TrackingPaper.tex

\begin{abstract}
  This paper describes the track-finding algorithm that is used for event reconstruction in the \BelleII experiment operating at the SuperKEKB \PB-factory in Tsukuba, Japan.
  The algorithm is designed to balance the requirements of a high efficiency
  to find charged particles with a good track parameter resolution,
  a low rate of spurious tracks, and a reasonable demand on CPU resources.
  The software is implemented in a flexible, modular manner and employs a diverse selection of global and local track-finding algorithms to achieve an optimal performance.
\end{abstract}

\end{frontmatter}

% !TeX encoding = UTF-8
% !TeX spellcheck = en_US
% !TeX root = ./TrackingPaper.tex

\section{Introduction} \label{sec:introduction}

The SuperKEKB accelerator complex~\cite{Ohnishi2013} located at Tsukuba, Japan  is designed to achieve a world-record instantaneous luminosity for $\Ppositron\Pelectron$~collisions of \SI{8d35}{\per\square\centi\metre\per\second}.
The collisions of \SI{4}{\giga\eV} positron and \SI{7}{\giga\eV} electron beams are recorded by the upgraded successor of the Belle detector~\cite{Brodzicka2012}, which is called  \BelleII~\cite{Abe2010}.
The expected data sample with an integrated luminosity of \SI{50}{\per\atto\barn} will allow the \BelleII experiment to study \PB{} meson decays with unprecedented accuracy.

The high instantaneous luminosity poses, however, several additional challenges.
The signal and background rates are expected to increase significantly compared to those observed at Belle.
The larger data samples will act to reduce statistical uncertainties, this emphasizes the need to keep systematic effects under control.
The experiment therefore requires highly performing track-finding software, capable to cope with high rates and significant background,
while maintaining  high efficiency and resolution for particles with momenta as low as \SI[per-mode=symbol]{50}{\MeV\per c}.

The track-finding algorithms used in \BelleII are built on the experiment's modular software framework~\cite{Kuhr2019} and can be combined
for an optimal overall performance. The algorithms use both local and global track-finding methods based
on cellular automaton~\cite{Gardner1970au,glazov:1993au,Abt:2002he,Funke:2014dga} and
Legendre transformation~\cite{Alexopoulos2008}, respectively, as well as combinatorial
Kalman filter (CKF) approaches~\cite{Mankel1999, Mankel2004, CMSCollaboration2014, Aaboud}.
A specific feature of the \BelleII tracking is a heavy use of multivariate methods, based on the gradient boosted decision tree implementation provided by the FastBDT package~\cite{keck2016},
to improve background filtering and track-candidate search.
The performance of the track finding is estimated using a detailed simulation of the \BelleII detector using \HepProcess{\PUpsilonFourS \to \PB\APB} events with expected background overlaid.

The paper is organized as follows. \cref{sec:detector} describes the main components of the \BelleII tracking devices: the silicon-based vertex detector (VXD) and the central drift chamber (CDC).
Properties of signal events and background are discussed in \cref{sec:event}.
\cref{sec:simulation} describes the event simulation and methods used to gauge the tracking performance.
The description of the reconstruction of hits in each of the tracking detectors is given next in \cref{sec:input}.
The general strategy for track reconstruction is outlined in \cref{sec:high_level_description} after which the CDC track finding is explained in \cref{sec:cdc_algorithm}.
Track finding with the silicon vertex detector (SVD) using the concept of dedicated \SectorMaps{} and a local track finding algorithm is discussed in \cref{sec:vxd_algorithm} followed by the description of the CKF in \cref{sec:ckf_description}.
\cref{sec:track_finding_performance} presents performance studies of the \BelleII track finding using simulated events and \cref{sec:summary} summarizes the results.

% !TeX encoding = UTF-8
% !TeX spellcheck = en_US
% !TeX root = ./TrackingPaper.tex

\section{\BelleII Tracking System} \label{sec:detector}

The trajectories of the charged long-lived decay products of the \PB{}~mesons are measured by the \BelleII tracking detectors:
the silicon based vertex detector  and the central drift chamber.
The origin of most of these trajectories is in the proximity of the interaction point (IP).
The trajectories pass through the beam pipe which is comprised of two thin walls of beryllium enclosing a duct through which liquid paraffin flows.
The inner wall of the beam pipe is sputtered with a thin layer of gold to shield the VXD from synchrotron radiation.
The beam pipe radiation length for particles crossing it at a \SI{90}{\degree}~angle is \percentageNumber{0.79}.
A thin superconducting solenoid provides a magnetic field of about \SI{1.5}{\tesla} directed along the nominal mechanical axis of the CDC support cylinder.
A system of final focusing quadrupole  and compensating solenoid magnets is situated close to the IP.
The field remains fairly homogeneous and varies on the order of \percentageNumber{1} in the entire tracking volume.

In spherical coordinates, with the $z$~axis parallel to the CDC axis of symmetry and directed along the boost direction,
the CDC covers the $\theta$~range comprised between 17\degree\, and 150\degree\, and the full $\phi$~range.
Just outside the CDC there are additional detectors for the reconstruction of neutral particles and particle identification.

\begin{figure}[t]
  \centering
    \includegraphics[width=0.6\textwidth]{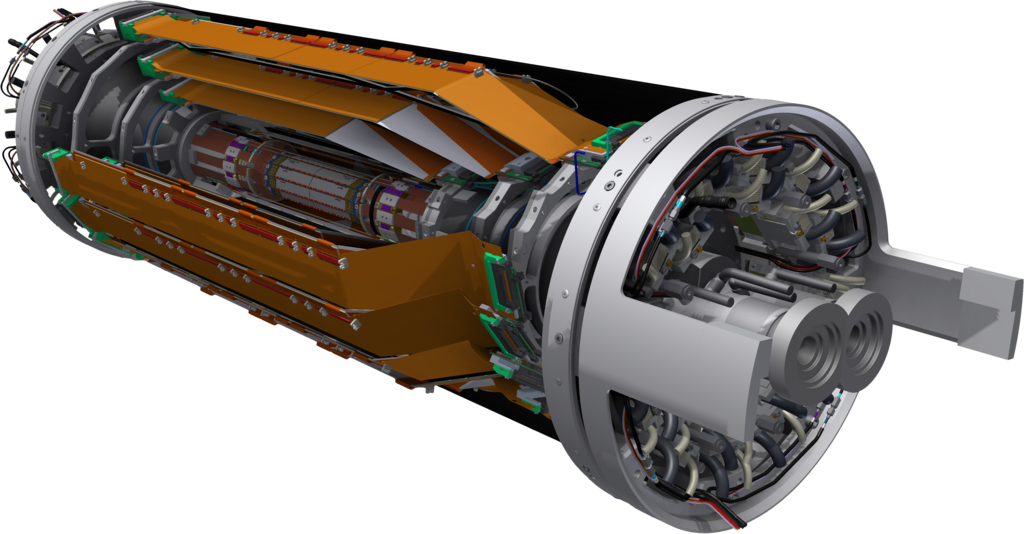}
    \caption{
      The \BelleII Vertex Detector volume.
      The four outer layers are the silicon vertex detector and the pixel detector is in the center.
    }
  \label{fig:VXD}
\end{figure}

\begin{table}[h]
  \caption{Specifications of the \BelleII PXD.}
  \label{tab:PXDProperties}
  \centering
  \begin{tabular}{cccccc} \hline
    Layer	& Radius	& Ladders	& Sensors	& Pixels$/$Sensor		  & Pitch                 						\\
		& (\SI{}{\mm})	&		&		& $u$ $\,\times\,$ $v$		& $u$ $\times$ $v$ (\SI{}{\micro\meter} $\times$ \SI{}{\micro\meter}) 	\\ \hline
    1		& 14		& 8		& 16		& 250$\times$768 		& 50 $\times$ (55 to 60) 						\\
    2		& 22		& 12		& 24		& 250$\times$768 		& 50 $\times$ (70 to 85)						\\ \hline
    Sum		&		& 20		& 40		& \num{7680000}			& 									\\ \hline
  \end{tabular}
\end{table}

A rendering of the VXD is shown in \cref{fig:VXD}.
The VXD is composed of two detectors, the  pixel detector (PXD) and the SVD,
which are based on DEPFET~\cite{Kemmer1987} and double-sided silicon strip technologies, respectively.
An overview of the key figures of the PXD is shown in \cref{tab:PXDProperties}.
The PXD consists of two approximately cylindrical layers with radii of \SI{14}{} and \SI{22}{\mm}.
The inner (outer) layer contains eight (twelve) ladders with a size of approximately \SI{1.5}{} by \SI{10}{\cm} (\SI{1.5}{} by \SI{13}{\cm}).
Each ladder is built by gluing two DEPFET modules together at their short edge.
In total there are 40 PXD sensors.
The ladders overlap with each other in $r$\nobreakdash-$\phi$ (local $u$~coordinate),
while there is a \SI{0.85}{\mm} gap between the two sensors on each ladder in $z$ (local $v$~coordinate).
The sensitive region of the PXD is \SI{75}{\micro\meter} in thickness while the edges,
which provide the mechanical stiffness to the structure and make the PXD ladder self-supporting, are \SI{450}{\micro\meter} thick.
One of the two long sides of each ladder is equipped with twelve switchers, six for each module.
These switchers are the only PXD ASICs (Application Specific Integrated Circuits) inside the tracking volume.
The other ASICs of the PXD are on the two short edges of each ladder in close contact to the cooling blocks that support the detector.
The structure is extremely light with the equivalent thickness for a PXD layer of \percentageNumber{0.2} of the radiation length.
In both layers, the PXD pixel matrix is organized in rows comprising of 250 pixels with a pitch of \SI{50}{\micro\meter} that run in the $u$~direction,
and columns comprising of 768 pixels with pitches varying between \SI{55}{\micro\meter} and \SI{85}{\micro\meter} that run along the $v$~direction.
In total the PXD comprises approximately eight million pixels.

\begin{table}[htbp]
  \caption{Specifications of the \BelleII SVD.}
  \label{tab:SVDProperties}
  \centering
  \begin{tabular}{cccccc} \hline
    Layer & Radius              & Ladders & Sensors & Strips$/$Sensor         & Pitch \\
	  & (\SI{}{\mm})	&         &         &  $u,\, v$                 & $u,\, v$ (\SI{}{\micro\meter}, \SI{}{\micro\meter}) \\ \hline
    3     & 39     	        & 7       & 14      & 768, 768                  & 50, 160 \\
    4     & 80     	        & 10      & 30      & 768, 512                  & 75 to 50, 240 \\
    5     & 104                 & 12      & 48      & 768, 512                  & 75 to 50, 240 \\
    6     & 135   	        & 16      & 80      & 768, 512                  & 75 to 50, 240 \\ \hline
    Sum   &        	        & 35      & 172     & \num{132096}, \num{91648}             &   \\ \hline
  \end{tabular}
\end{table}

\cref{tab:SVDProperties} shows the key figures of the SVD.
The SVD consists of four layers of double-sided silicon strip detectors.
All the layers have a barrel-shaped part with rectangular sensors.
The forward section of the  outermost three layers has a lamp-shade geometry made of trapezoidal sensors.
This setup minimizes the amount of material for the particles originating from the IP.
The radii of the four SVD layers range from \SI{39}{\mm} to \SI{135}{\mm}.
The layers consist of 7 to 16 ladders, with 2 to 5 sensors per ladder, respectively.
Similarly to the PXD, the SVD ladders overlap in $u$ while there is a \SI{2}{\mm} gap between the sensors on each ladder in $v$.
Each sensor of the first layer of the SVD has 768 strips per side, with readout pitches of \SI{50}{\micro\meter} on the side measuring the $u$ coordinate
and \SI{160}{\micro\meter} on the side measuring the $v$ coordinate.
The barrel sensors of the three outer layers have 768 strips with a readout pitch of \SI{75}{\micro\meter} in $u$ and 512 strips with a readout pitch of \SI{240}{\micro\meter} in $v$.
The slanted sensors of these layers have the same number of strips in the respective directions, and the same pitch in $v$.
The pitch in $u$-direction varies from \SI{75}{\micro \meter} at the back to \SI{50}{\micro \meter} at the front side, due to the trapezoidal shape.
The readout strips are interleaved with floating strips to improve the spatial resolution.
In total, there are 172 SVD sensors with about 220 thousand read-out strips.
Each SVD sensor has a thickness of \SI{320}{\micro\meter}.
The contribution to the overall radiation length due to mechanical support structure,
electronic read-out and cooling is kept at a minimum so that the material of the outer SVD layers is equivalent to \percentageNumber{0.6} radiation length at normal incidence.

\begin{figure}[htbp]
  \centering
    \includegraphics[width=0.35\textwidth]{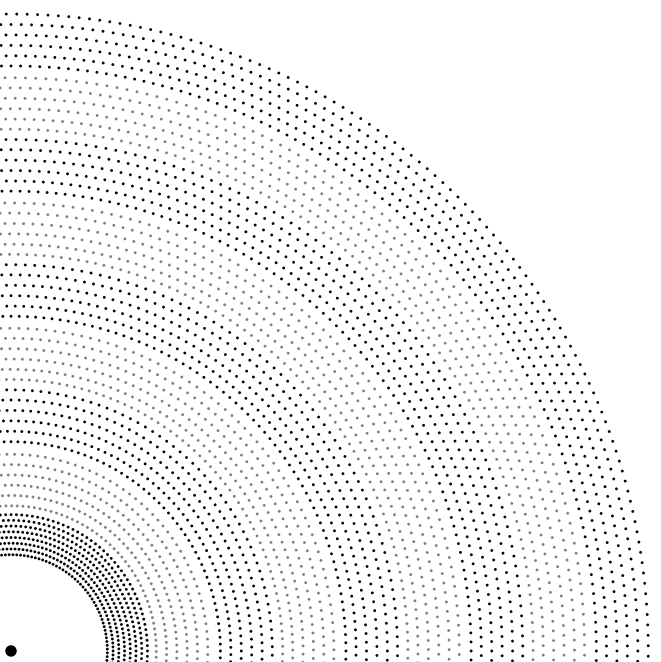}
    \includegraphics[width=0.55\textwidth]{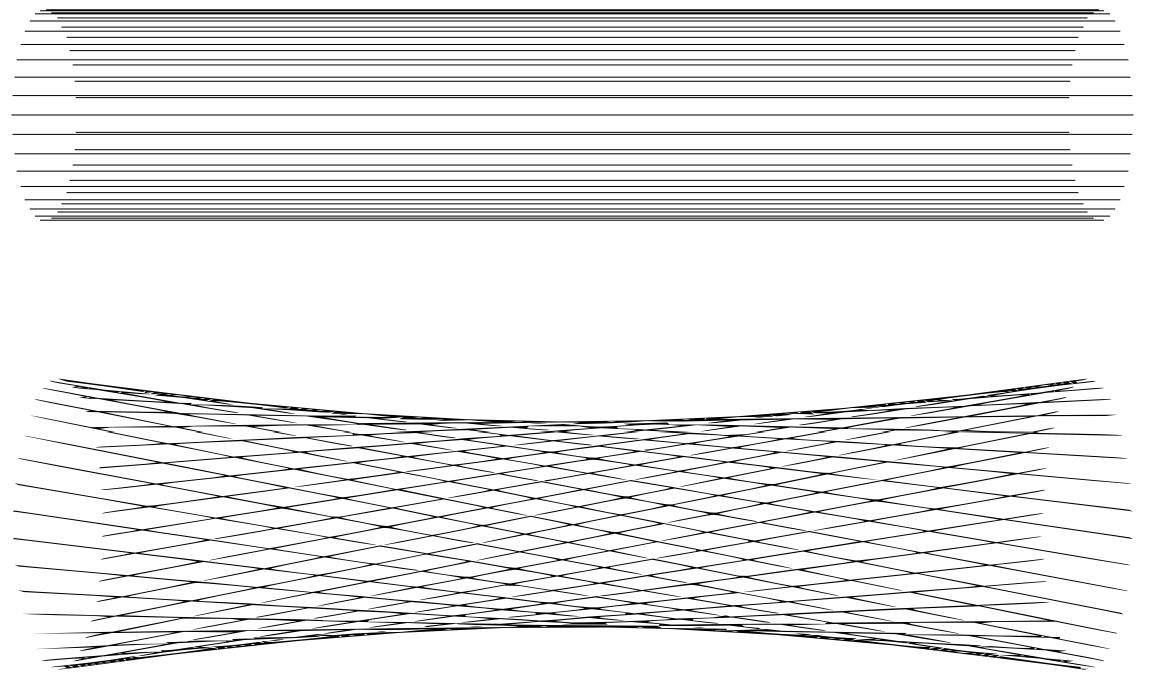}
  \caption{
    Left: A quadrant of a slice of the $r$\nobreakdash-$\phi$~projection of the drift chamber.
    The innermost superlayer contains eight layers, all others contain six.
    Right: A visualization of stereo wires (bottom) relative to axial wires (top).
    The skew is exaggerated.
  }
  \label{fig:Skew}
\end{figure}

The main specifications of CDC are given in \cref{tab:cdc}.
The inner volume of the CDC contains about \SI{50000}{} sense and field wires, defining drift cells with a size of about \SI{2}{\cm}.
The electric field in the drift cells is approximately cylindrical leading to a two-fold ambiguity with the same drift time measured for the tracks passing at the same distance on either side of the sense wire (\emph{left-right passage} ambiguity).
The sense wires are arranged in layers, where six or eight adjacent layers are combined in a superlayer, as seen in \cref{fig:Skew}.
The outer eight superlayers consist of six layers with 160 to 384 wires.
The innermost superlayer has eight layers with 160 wires in smaller (half-size) drift cells to cope with the increasing background towards smaller radii.
The superlayers alternate between axial (A) orientation, aligned with the solenoidal magnetic field, and stereo (U, V) orientation.
Stereo wires are skewed by an angle between \SI{45.4}{} and \SI{74}{\milli\radian} in the positive and negative direction.
The direction changes sign between U and V layers, with a total superlayer configuration of AUAVAUAVA.
The drift distance resolution of the drift chamber is about \SI{120}{\micro\meter}.
By combining the information of axial and stereo wires it is possible to reconstruct a full three-dimensional trajectory.

\begin{table}[tb]
  \caption{\label{tab:cdc}Specification of the \BelleII CDC.}
  \centering
  \begin{tabular}{cccccc}
    \hline
    Layer    & Radius of       & Number     & Drift               & Average \\
             & Sense Wires     & of Wires   & Cell Size           & Resolution \\
             & (\SI{}{\mm})    &            & (\SI{}{\cm})        & (\SI{}{\micro\meter}) \\ \hline
    1 to 56  &  168 to 1111.4  & 160 to 384 & $\sim$1 to $\sim$2  & 120 \\ \hline
  \end{tabular}
\end{table}

% !TeX encoding = UTF-8
% !TeX spellcheck = en_US
% !TeX root = ./TrackingPaper.tex

\section{\BelleII Events and Background} \label{sec:event}

The events recorded by the \BelleII experiment can be classified according to the $\Ppositron{}\Pelectron$~scattering process occurring at the interaction point.
The main category is composed of the \PUpsilonFourS{}~events in which the annihilation of an electron-positron pair produces an \PUpsilonFourS{}~resonance.
This resonance decays promptly  into a quantum entangled state of two \PB{}~mesons.
The \PB{}~meson decay vertices have an average spatial separation of $\sim$\SI{130}{\um}.
Thus a tracking detector resolution significantly better than that is required to resolve them.
This is crucial for the measurements of the time dependent CP and T violation as well as tests of the CPT symmetry in the \PB{}~meson system.
The decay-vertex resolution relies on the spatial resolution of the PXD sensors as well as their proximity to the IP 
in order to reduce the extrapolation lever arm,
and thus the effects of multiple Coulomb scattering on the measurement of the impact parameters.

\begin{figure}
  \centering
  \includegraphics[width=0.75\textwidth]{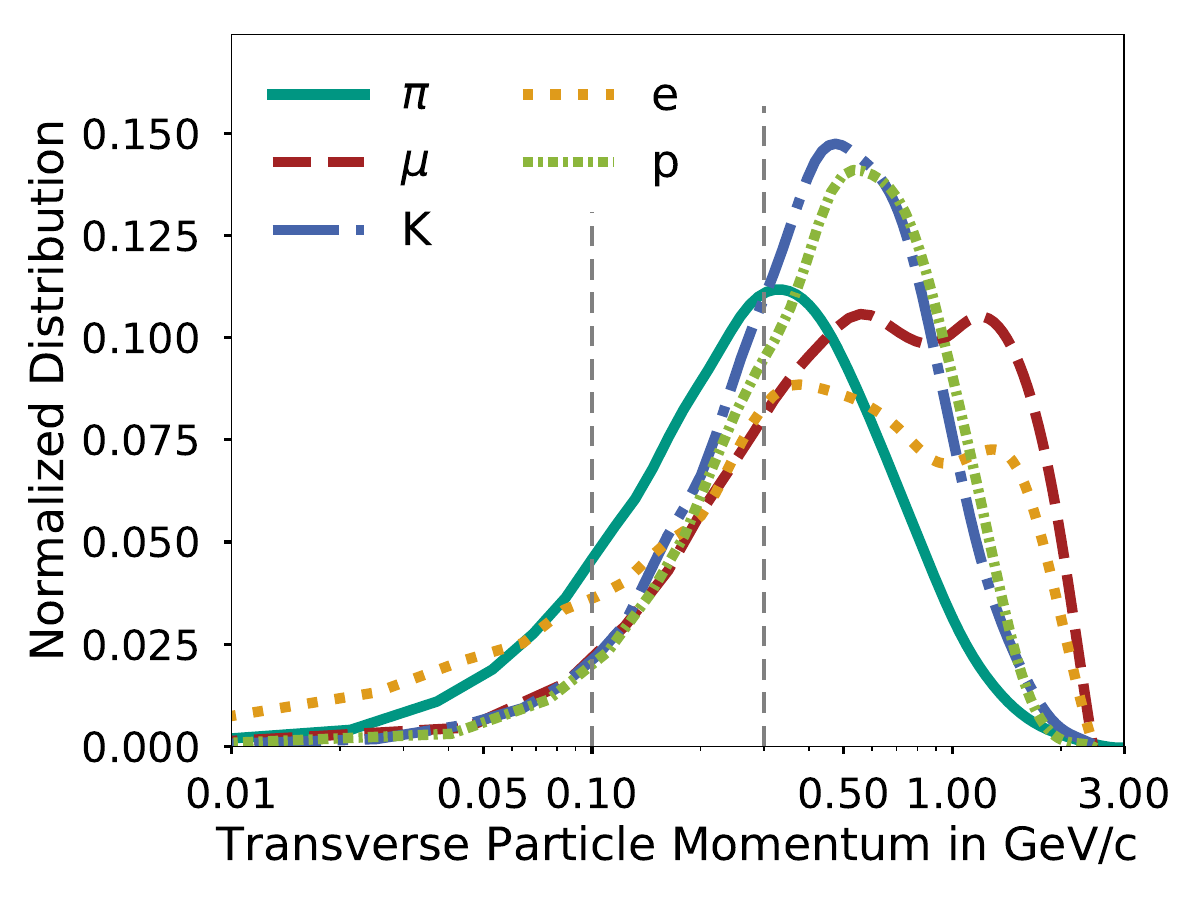}
  \caption{
    Transverse momentum distributions of primary charged particles as simulated for \PUpsilonFourS{}~events.
    A logarithmic scale is used for the $x$~axis.
    The distribution of each charged particle type is normalized to the total number of tracks from the respective type.
    The vertical line at \SI[per-mode=symbol]{100}{\MeV\per c} indicates the transverse momentum threshold below which a track can only be found by the SVD.
    Charged particles with transverse momenta below the value \SI[per-mode=symbol]{300}{\MeV\per c} marked by the second vertical line can curl inside the CDC volume.
  }%
  \label{fig:pt_distribution}%
\end{figure}

Studies of (semi)leptonic \PB{}~decays often require the reconstruction of the missing neutrino by exploiting \FourMomentum conservation.
Hence the tracking algorithm needs to find all of the charged final state particles.
This is demanding since there are about $11$~tracks per event on average.
Moreover, the momentum spectrum of the particles is quite soft, ranging from a few tens of \SI[per-mode=symbol]{}{\MeV\per c} to a few \SI[per-mode=symbol]{}{\GeV\per c} (\cref{fig:pt_distribution}).
It is also essential to keep the rate of fake and duplicate tracks as low as possible.

\begin{figure}
  \centering
  \begin{tikzpicture}[fading style/.style={preaction={fill=#1,opacity=.8,
                   path fading=circle with fuzzy edge 20 percent}}]
                   
  \newcommand{\radaa}{-30.0}% {0.0}
  \newcommand{\radab}{232.08}% {262.08}
  \newcommand{\radba}{\radab}
  \newcommand{\radbb}{285.72}% {315.72}
  \newcommand{\radca}{\radbb}
  \newcommand{\radcb}{306.60}% {336.60}
  \newcommand{\radda}{\radcb}
  \newcommand{\raddb}{323.52}% {353.52}
  \newcommand{\radea}{\raddb}
  \newcommand{\radeb}{330.00}% {360.00}
  
  \newcommand{\radamarker}{130.0}
  \newcommand{\radbmarker}{257.0}
  \newcommand{\radcmarker}{296.0}
  \newcommand{\raddmarker}{315.0}
  \newcommand{\raddmarkerb}{-2.5}
  \newcommand{\rademarker}{327.0}
  \newcommand{\rademarkerb}{30.0}
  
  \newcommand{\particlemarkersize}{0.3mm}
  
  \newcommand{\innerradius}{0.3cm}
  \newcommand{\outerradius}{1.0cm}
  
  \newcommand{\deptheffect}{1.0mm}
  \newcommand{\shadoweffect}{2}
  \definecolor{shadow_effect_color}{HTML}{000000}
  
  \definecolor{pion_color}{HTML}{009682}
  \definecolor{kaon_color}{HTML}{4664aa}
  \definecolor{electron_color}{RGB}{223, 155, 27}
  \definecolor{muon_color}{HTML}{A22223}
  \definecolor{proton_color}{HTML}{8cb63c}

  \begin{scope}[xscale=3.5,yscale=2.5]
    \path[fading style=black,transform canvas={yshift=-40pt}] (0,0) circle (\outerradius);
    \fill[gray](0,0) circle (\innerradius);
    \path[fading style=white,transform canvas={yshift=-16mm}] (0,0) circle (0.65cm);
    \draw[yshift=-\deptheffect](0,0) circle (\innerradius);

    % Inner shadows
    \shadedraw[top color=pion_color,,bottom color=pion_color!\shadoweffect!shadow_effect_color,draw=black,very thin]
        (\radab:\innerradius)--++(0,-\deptheffect) arc(\radab:\radaa:\innerradius)--++(0,\deptheffect)  arc(\radaa:\radab:\innerradius)--cycle;

    % Top ring
    \begin{scope}[draw=black,thin]
       \fill[pion_color](\radaa:\innerradius)--(\radaa:\outerradius) arc(\radaa:\radab:\outerradius)--(\radab:\innerradius) arc(\radab:\radaa:\innerradius);
       \fill[kaon_color](\radba:\innerradius)--(\radba:\outerradius) arc(\radba:\radbb:\outerradius)--(\radbb:\innerradius) arc(\radbb:\radba:\innerradius);
       \fill[electron_color](\radca:\innerradius)--(\radca:\outerradius) arc(\radca:\radcb:\outerradius)--(\radcb:\innerradius) arc(\radcb:\radca:\innerradius);
       \fill[muon_color](\radda:\innerradius)--(\radda:\outerradius) arc(\radda:\raddb:\outerradius)--(\raddb:\innerradius) arc(\raddb:\radda:\innerradius);
       \fill[proton_color](\radea:\innerradius)--(\radea:\outerradius) arc(\radea:\radeb:\outerradius)--(\radeb:\innerradius) arc(\radeb:\radea:\innerradius);
    \end{scope}
    \draw[thin,black](0,0) circle (\innerradius);

    % Outer shadows
    \shadedraw[bottom color=pion_color,top color=pion_color!\shadoweffect!shadow_effect_color,draw=black,very thin]
    (180.0:\outerradius) -- ++(0,-\deptheffect) arc (180.0:\radab:\outerradius) -- ++(0,\deptheffect) arc (\radab:180.0:\outerradius) -- cycle;
    \shadedraw[bottom color=kaon_color,top color=kaon_color!\shadoweffect!shadow_effect_color,draw=black,very thin]
    (\radba:\outerradius) -- ++(0,-\deptheffect) arc (\radba:\radbb:\outerradius) -- ++(0,\deptheffect) arc (\radbb:\radba:\outerradius) -- cycle;
    \shadedraw[bottom color=electron_color,top color=electron_color!\shadoweffect!shadow_effect_color,draw=black,very thin]
    (\radca:\outerradius) -- ++(0,-\deptheffect) arc (\radca:\radcb:\outerradius) -- ++(0,\deptheffect) arc (\radcb:\radca:\outerradius) -- cycle;
    \shadedraw[bottom color=muon_color,top color=muon_color!\shadoweffect!shadow_effect_color,draw=black,very thin]
    (\radda:\outerradius) -- ++(0,-\deptheffect) arc (\radda:\raddb:\outerradius) -- ++(0,\deptheffect) arc (\raddb:\radda:\outerradius) -- cycle;
    \shadedraw[bottom color=proton_color,top color=proton_color!\shadoweffect!shadow_effect_color,draw=black,very thin]
    (\radea:\outerradius) -- ++(0,-\deptheffect) arc (\radea:\radeb:\outerradius) -- ++(0,\deptheffect) arc (\radeb:\radea:\outerradius) -- cycle;
    \shadedraw[bottom color=pion_color,top color=pion_color!\shadoweffect!shadow_effect_color,draw=black,very thin]
    (\radaa:\outerradius) -- ++(0,-\deptheffect) arc (\radaa:0.0:\outerradius) -- ++(0,\deptheffect) arc (0.0:\radaa:\outerradius) -- cycle;

    % Top border lines
    \draw[very thin]
	  (\radab:\innerradius) -- (\radab:\outerradius)
	  (\radaa:\innerradius) -- (\radaa:\outerradius)
	  (\radbb:\innerradius) -- (\radbb:\outerradius)
	  (\radcb:\innerradius) -- (\radcb:\outerradius)
	  (\raddb:\innerradius) -- (\raddb:\outerradius)
	  (0,0) circle (\outerradius)
	  (90:\innerradius) arc (90 :135:\innerradius);

    % Coordinate definitions
    \coordinate (right border) at (1.5cm,0cm);
    \coordinate (left border) at (-1.5cm,0cm);
    
    \coordinate (l1) at (\radamarker:0.75 cm);
    \coordinate (l2) at (\radbmarker:0.75 cm);
    \coordinate (l3) at (\radcmarker:0.8 cm);
    \coordinate (l4) at (\raddmarker:0.8 cm);
    \coordinate (l4b) at (\raddmarkerb:1.2 cm);
    \coordinate (l5) at (\rademarker:0.75 cm);
    \coordinate (l5b) at (\rademarkerb:1.38 cm);

    % Markers
    \begin{scope}[lab/.style={gray!50!black,thick,draw}]
       \fill[lab] (l1) circle(\particlemarkersize) -- (l1-| left border) node[anchor=south west] {\Ppipm{}}
                                                           node[anchor=north west] {\percentageNumber{72.8}};  % 0.0 --> 262.08 degrees
       \fill[lab] (l2) circle(\particlemarkersize) -- (l2-| left border) node[anchor=south west] {\PKpm{}}
                                                           node[anchor=north west] {\percentageNumber{14.9}};  % 262.08 --> 315.72 degrees
       \fill[lab] (l3) circle(\particlemarkersize) -- (l3-| right border) node[anchor=south east] {\Pepm{}}
                                                           node[anchor=north east] {\percentageNumber{5.8}};  % 315.72 --> 336.60 degrees
       \fill[lab] (l4) circle(\particlemarkersize) -- (l4b);
       \fill[lab] (l4b) -- (l4b-| right border) node[anchor=south east] {\Pmupm{}}
                                                           node[anchor=north east] {\percentageNumber{4.7}};  % 336.60 --> 353.52 degrees
       \fill[lab] (l5) circle(\particlemarkersize) -- (l5b);
       \fill[lab] (l5b) -- (l5b-| right border) node[anchor=south east] {\HepParticle{\Pp}{}{\pm}}
                                                           node[anchor=north east] {\percentageNumber{1.8}};  % 353.52 --> 360.00 degrees
    \end{scope}
  \end{scope}
\end{tikzpicture}
  \caption{Fractions of charged particle types in generic \PUpsilonFourS{}~events.}
  \label{fig:particles_pie_chart}
\end{figure}
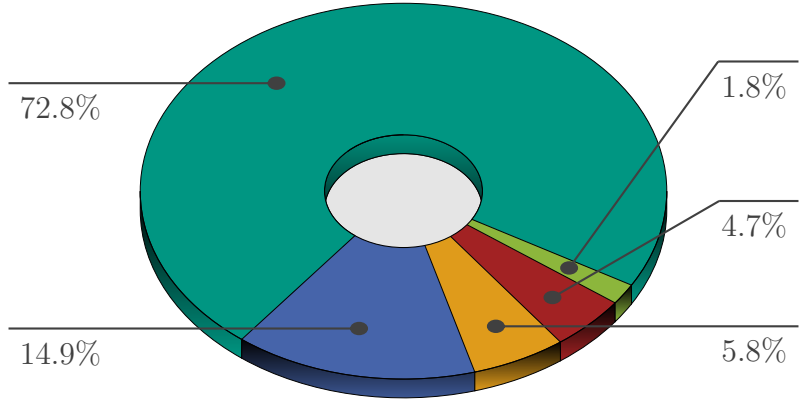

The reconstruction of particles with momenta below \SI[per-mode=symbol]{200}{\MeV\per c} is particularly challenging
since the trajectories are heavily affected by multiple Coulomb scattering and by energy loss in the material.
Moreover, only the measurements of the four layers of the SVD are available to the pattern recognition algorithms for most of the tracks in this low momentum region.
The soft momentum spectrum is also challenging for the CDC since particles with momenta below \SI[per-mode=symbol]{300}{\MeV\per c} can loop several times in the CDC volume producing hundreds of hits.
The relative abundance of the long lived charged particles produced in \PUpsilonFourS{}~decays is illustrated in \cref{fig:particles_pie_chart}.

Other categories of events are also of importance to the experiment.
Most notably \Ptau{}\nobreakdash-pair and \Pcharm{}\APcharm{} events improve the existing limits and measurements on the \Ptau{}~lepton sector and on the charmed mesons.
The experiment will also be used to search for non--Standard Model particles, i.e.\ dark photons, axion-like particles,
or magnetic monopoles that might be produced directly in $\Ppositron{}\Pelectron$~collisions.
These events are characterized by a lower track multiplicity, a stiffer momentum spectrum, and by a less spherical event topology.

Particles lost by beam-gas and Touschek scattering, as well as due to non-linearities of the machine lattice, lead to additional hits in the detector.
The occupancy due to this machine background is expected to be very high as a consequence of the high beam currents,
small emittances, and large beam-beam tune shifts needed to reach the design luminosity~\cite{Lewis:2018ayu}.
The electromagnetic processes occurring at the interaction point, radiative Bhabha and electron-positron pair production,
whose cross sections are of the order of several \SI{}{\milli\barn}, are going to be the leading effects for the beam particle loss rate at nominal luminosity.

The VXD occupancy is expected to be largely dominated by soft electron-positron pairs produced at the IP by the process \HepProcess{\Ppositron \Pelectron \to \Ppositron \Pelectron \Ppositron \Pelectron}.
The forward and backward sections of the SVD may be hit by an electromagnetic shower originating from Bhabha electrons interacting in the support structure of the final focusing magnets.
The number of background hits exceeds the signal hits by two orders of magnitude resulting in a PXD inner layer pixel occupancy close to \percentageNumber{2} and an SVD inner layer strip occupancy close to \percentageNumber{3}.

\begin{figure}
  \centering
  \includegraphics[width=0.6\textwidth]{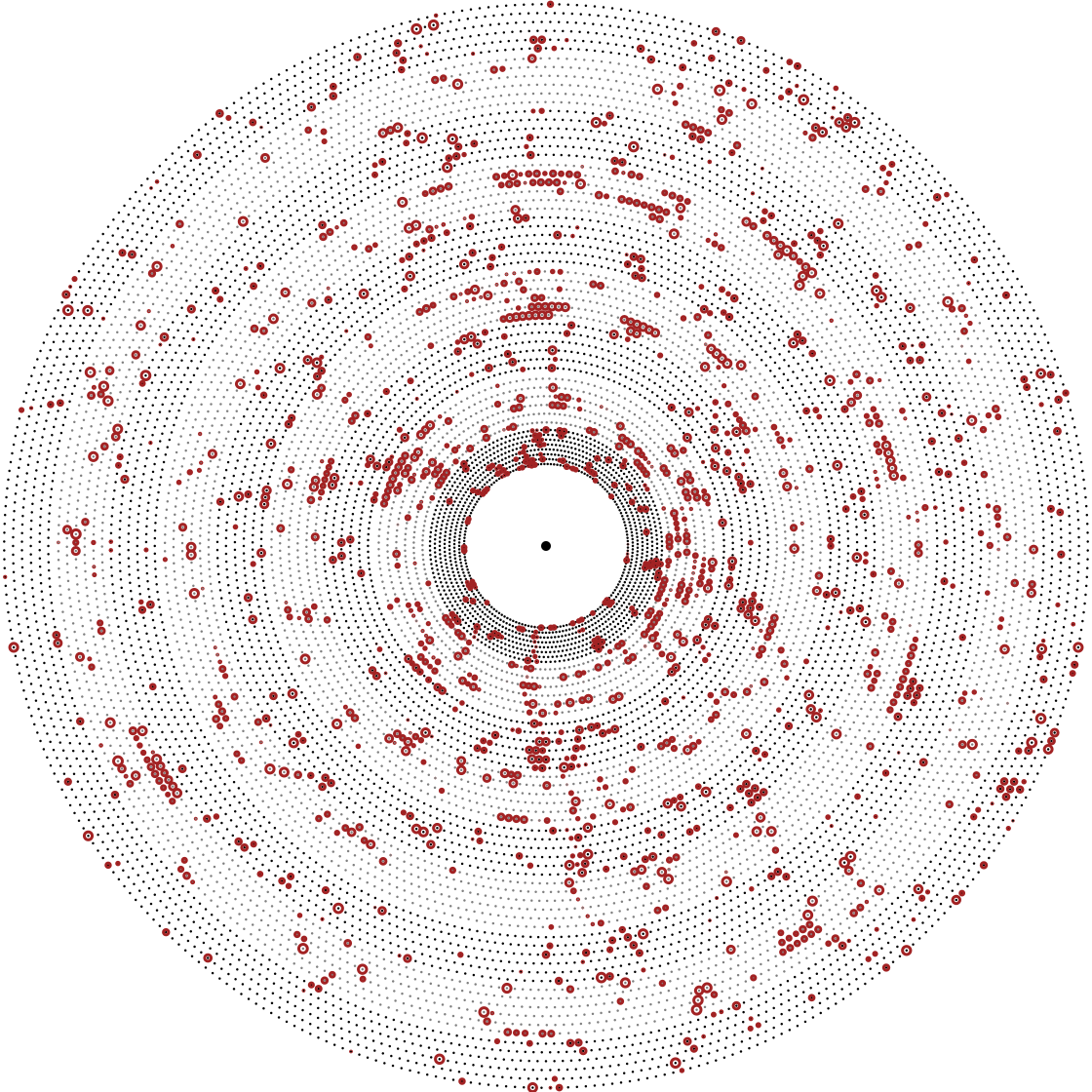}
  \caption{CDC measurements produced by simulated beam-induced background anticipated for the nominal instantaneous luminosity.}
  \label{fig:cdc-display}
\end{figure}

The CDC occupancy is also expected to be dominated by the hits left by particles coming from electromagnetic showers initiated by beam particles.
These interact with the material around the final focusing magnets which are well inside the CDC volume.
\cref{fig:cdc-display} shows the CDC measurements produced by simulated beam-induced background for the nominal instantaneous luminosity.

% !TeX encoding = UTF-8
% !TeX spellcheck = en_US
% !TeX root = ./TrackingPaper.tex

\section{Simulation and Track Finding Efficiency Definition} \label{sec:simulation}

\begin{figure}[]
  \centering
  \subfloat[]{
    \includegraphics[scale=0.355]{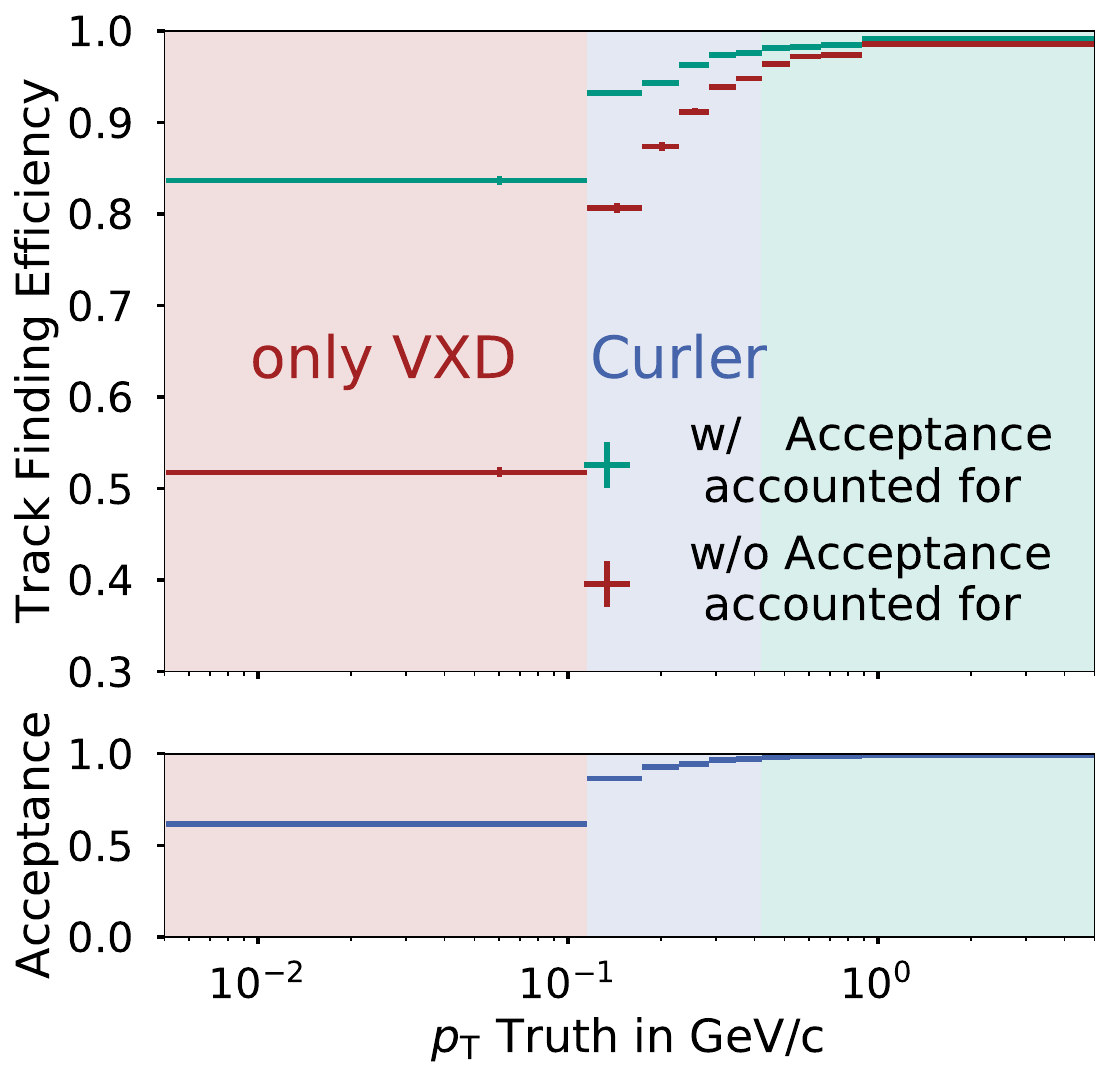}
  }
  \subfloat[]{
    \includegraphics[scale=0.355]{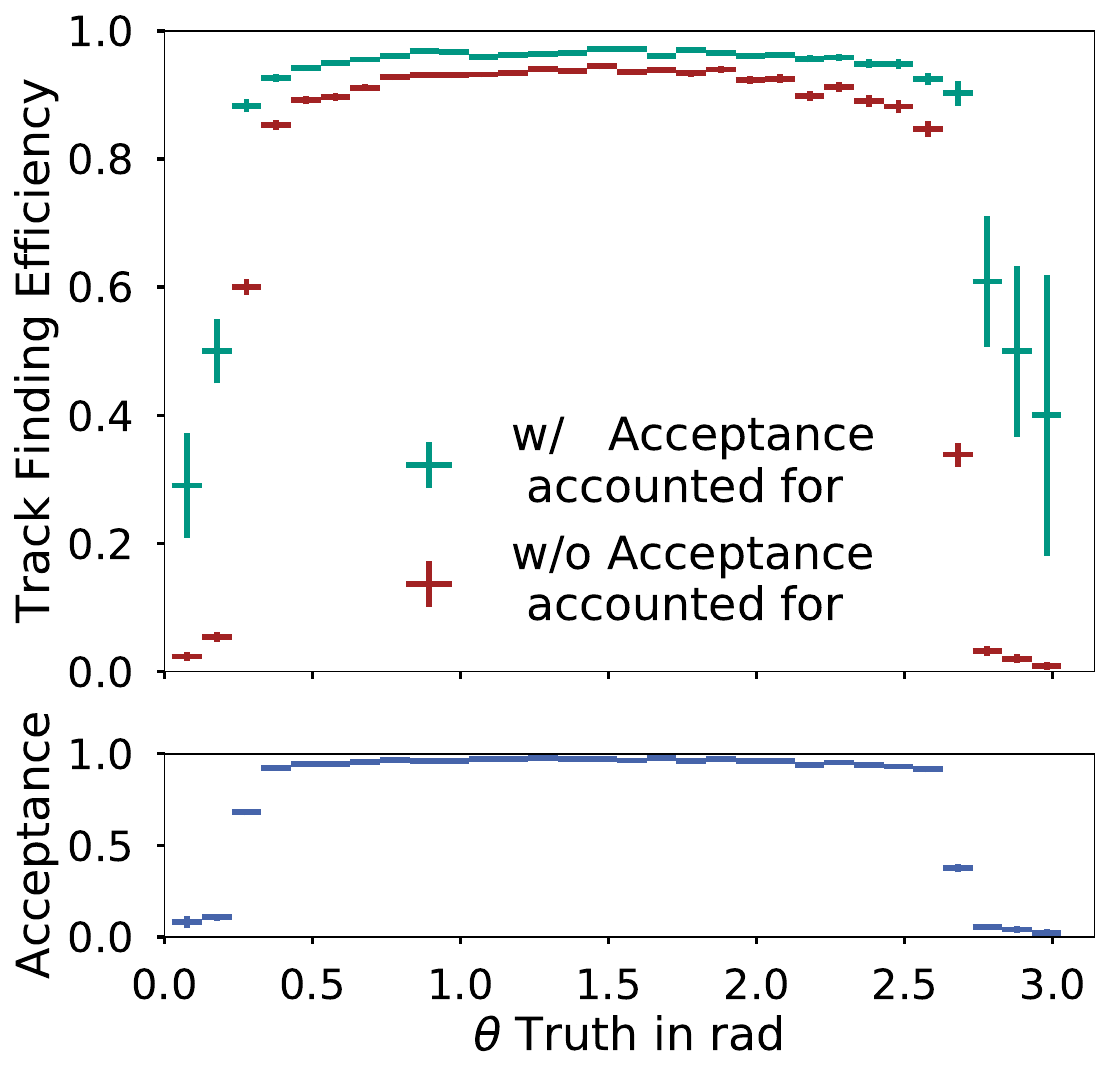}
  }
  \caption{
    Illustration of the effect of the detector acceptance on the track finding efficiency for primary particles dependent on
    (a) the true transverse momentum $p_\mathrm{T}$ and (b) the true polar angle $\theta$.
    For this purpose the track finding efficiency is shown normalized to the number of MC-tracks,
    where the acceptance is already accounted for in the denominator (labeled \textit{w/ Acceptance accounted for}),
    and normalized to all primary and stable, charged Monte Carlo particles,
    where the effect of the nonperfect acceptance results in a lower efficiency.
    The acceptance itself is shown as ratio of the two finding efficiency below both plots.
    In these figures, all tracks from \PUpsilonFourS{} decays
    are used for the evaluation of the acceptance and the efficiency.
  }
  \label{fig:simulation:acceptance}
\end{figure}

A full simulation tool based on Geant4~\cite{Agostinelli2003} is used to model the detector and collider properties.
Using the  information from the particle generator and the Geant4 simulation of the particles traversing the detector volume,
an ideal track finder, called Monte Carlo (MC) track finder, is implemented.
Its performance is limited only by the detector acceptance, efficiency and resolution, and by definition cannot be surpassed.
A set of figures of merit has been developed to qualify and tune the track finding algorithms.
The analysis is limited to tracks identified by the MC track finder (MC-tracks from now on)
having enough hits to completely determine the five parameters of the helix-like trajectory.

A good track finding algorithm should behave as closely as possible to the MC track finder.
In particular, each track should be assigned all of the hits of one and only one MC particle.
Two figures of merit are defined for each pair of MC-track and a track found by the pattern recognition (PR-track):

\begin{itemize}
\item The \emph{hit efficiency} quantifies how efficient the
  pattern recognition is in identifying \emph{all} the hits belonging to a single particle.
  It is defined as the fraction of hits of a given MC-track contained in a given PR-track.
  Ideally, there should be one and only one PR-track containing all the hits of a given MC-track,
  thus the \emph{hit efficiency} should be \percentageNumber{100} for the correct pair and zero for all others.
\item The \emph{hit purity} quantifies how precise the pattern recognition is in identifying the hits belonging to \emph{only one} particle.
  It is defined as the fraction of hits of a given PR-track contained in a given MC-track.
  Ideally, there should be one and only one MC-track to which all the hits of a given PR-track belong,
  thus the \emph{hit purity} should be \percentageNumber{100} for the correct pair and zero for all others.
\end{itemize}

A PR-track is defined as \emph{matched} to a given MC-track if the hit purity exceeds \percentageNumber{66} and the hit efficiency exceeds \percentageNumber{5}.
The low hit efficiency requirement accounts for low momentum tracks curling in the tracking volume which may leave several hundred of hits.

If there are two or more PR-tracks that are matched to the same MC-track,
the PR-track with the highest hit purity is defined as the correctly identified match and the remaining PR-tracks are defined as \emph{clones}.
If multiple PR-tracks have the same hit purity, the hit efficiency is used in addition to the purity to identify the match.

The \emph{track finding efficiency} is defined as the fraction of matched MC-tracks over all MC-tracks.
This definition of the track finding efficiency factors out the effect of the limited acceptance of the detector,
allowing for an easier interpretation of the performance of the algorithms.
\cref{fig:simulation:acceptance} shows the effect of the detector acceptance on the track finding efficiency
by comparing the fraction of matched MC-tracks normalized to all MC-tracks
and normalized to all primary charged Monte Carlo particles.
For both cases the same data set of \PUpsilonFourS{}~events with nominal background are used.
Additionally, the ratio of the two efficiencies, which corresponds to the detector acceptance, is shown.

If the PR-track fails the purity requirement, e.g.\ the PR-track is made up of hits from two MC-tracks,
each one with a hit purity below \percentageNumber{66} or the PR-track is made of background hits, it is defined as a \emph{fake}.

% !TeX encoding = UTF-8
% !TeX spellcheck = en_US
% !TeX root = ./TrackingPaper.tex

\section{Input to Tracking Algorithms} \label{sec:input}

\subsection*{PXD Reconstruction}
In order to reduce the \BelleII data rate to an acceptable level, events are required to pass a software-based high-level trigger (HLT).
Data from the PXD do not contribute to the HLT decision, and are therefore buffered in the readout chain.
In case an event is accepted, the track information from the HLT is used to define so-called \emph{Regions Of Interest} (ROIs) on the PXD planes.
Only PXD hits within these ROIs are stored.

Neighboring pixels with a charge above a threshold are combined into clusters.
The cluster position and charge are taken as input for the tracking algorithm.

\subsection*{SVD Reconstruction}

The SVD reconstruction software provides in addition to cluster charge and position information also cluster time information to the tracking algorithms.

SVD clusters are formed by combining adjacent strips with a signal-over-noise ratio (SNR) above three.
At least one strip in the cluster is required to have a SNR above five.
The charge of the cluster is computed as the sum of the charges of the strips,
while the cluster time and position are evaluated as the charge-weighted average of the strip times and positions, respectively.
The cluster position resolution depends on the cluster size and strip pitch, as shown in \cref{tab:svd}.

\begin{table}
  \caption{\label{tab:svd}
    Cluster position resolutions for different cluster sizes (one, two and larger than two) and strip pitch, evaluated on MC simulation.
    Resolutions are measured as \percentageNumber{68} coverage of the residual distributions.
  }
  \centering
  \begin{tabular}{c c c c c c}
    \hline
    Sensor & Side & Pitch & \multicolumn{3}{ c }{Resolution (\SI{}{\micro\meter})} \\
	   &      & (\SI{}{\micro\meter}) & Size = 1 & Size = 2 & Size $> 2$\\
    \hline
    \multirow{2}{*}{ Layer 3} & $u$ & 50  & 5.2  &  3.7  & 7.6 \\
			      & $v$ & 160 & 18.1 & 12.1  & 18.0 \\
    \hline
    \multirow{2}{*}{Slanted}  & $u$ & \numrange[range-phrase = --]{52}{75}   & 6.8  & 4.5  & 8.6 \\
			      & $v$ & 240  & 34.4 & 18.0 & 21.4 \\
    \hline
    \multirow{2}{*}{Barrel}  & $u$ & 75   & 7.7  & 5.1  & 8.8 \\
                              & $v$ & 240  & 24.8 & 17.1 & 20.5 \\
    \hline
  \end{tabular}
\end{table}

The creation of \SpacePoints{} follows the clustering and is achieved by combining all clusters on one side of a sensor with clusters on the other side.
The only requirement is that the cluster time (on both sides) is greater than a minimum value.
The cluster time information helps to reject the majority of the out-of-time clusters, created by beam-background particles produced before, or after the collision event of interest.
The SVD cluster time resolution varies between \SI{2}{\nano\second} (for Bhabha events) and \SI{4}{\nano\second} (hadronic events), being slightly better on the $v$ side due to a faster response of the electronics on that side.
The good time resolution allows the algorithm to reject \percentageNumber{60} of the background \SpacePoints{}, while retaining \percentageNumber{100} of the interesting ones.
The cluster time information provided by the SVD is also used later in the reconstruction, in the pattern recognition step, as described in \cref{sec:vxd_algorithm}.

\subsection*{CDC Reconstruction}

The front-end read-out electronics of the CDC use a time-to-digital converter with a \SI{1}{\nano\second} resolution.
This is used to measure the time between the event's trigger signal and the arrival of the drift electrons at the sense wire, the so-called \emph{drift time}.
With the $x$\nobreakdash-$t$~relation function, which is an approximation between drift time and distance parameterized in various areas of the CDC,
the actual relative distance between the sense wire and a passing particle can be computed and used for track finding purposes.
An additional front-end read-out provides amplitude information, sampled at \SI{33}{\MHz}.
This information is used for the determination of the energy loss, employed by the particle identification.
It can also be used to separate signal and background hits.

% !TeX encoding = UTF-8
% !TeX spellcheck = en_US
% !TeX root = ./TrackingPaper.tex

\usetikzlibrary{arrows, arrows.meta, decorations.markings, shapes, positioning, backgrounds, calc, fit, shapes.geometric}
\tikzstyle{pathelement} = [rectangle, draw, text width=10em, minimum height=1.5em, text centered, line width=0.3mm]
\tikzstyle{vecArrow}[black] = [
    #1,
    thick, 
    double distance=1.4pt, 
    shorten <= -0.85pt,
    arrows={-Triangle[angle=60:7pt,#1,fill=white]},
    postaction = {draw, -, line width = 1.4pt, white, shorten >= 4.0pt, shorten <= 4.0pt}
  ]
\definecolor{chapterblue}{HTML}{4664aa}
\definecolor{chapterred}{HTML}{A22223}
\definecolor{chaptergreen}{HTML}{009682}
\definecolor{chapterorange}{RGB}{223, 155, 27}
\definecolor{chapterlightblue}{HTML}{23a1e0}

\section{High-Level Description of the Tracking Setup} \label{sec:high_level_description}
\begin{figure}[h]
  \centering
  \begin{tikzpicture}
    \definecolor{colorip}{rgb}{0.5,0.5,0.5};
    \begin{axis}[
      unit vector ratio*=1 1 1,
      xmin=-50,
      xmax=55,
      ymin=-20,
      ymax=100,
      grid=major,
      grid style={dashed, gray!30},
      axis y line=center,
      axis x line=middle,
      xlabel=$x$,ylabel=$y$,
      every axis x label/.style={at={(ticklabel* cs:1.01)}, anchor=west,},
      scale=0.7
      ]
      %ip and the center of the circle
      \fill[colorip] (axis cs:0,0) ellipse [x radius=2, y radius=2];
      \fill[chapterred] (axis cs:-28,28) ellipse [x radius=2, y radius=2];
      %trajectory
      \draw[very thick] (axis cs:10,-10) arc [x radius=55, y radius=55, start angle=-45, end angle = 100];
      %arrows on the circle
      \draw[thick, ->] (axis cs:10.5,67.3) -- (axis cs:10,67.8);
      %d_0
      \draw[very thick, <->, chapterblue] (axis cs:0.05,0.0) -- (axis cs:10,-10) node [pos=0.4, below right] {$d_0$};
      %guide lines
      \addplot[domain=-50:55] {x-20};
      \addplot[domain=0:55] {0};
      %direction of propagation
      \draw[very thick, ->, chapterblue] (axis cs:10,-10) -- (axis cs:30,10);
      %radius
      \draw[very thick, ->, chapterred] (axis cs:-28,28) -- (axis cs:22,50) node [pos=0.4, below] {$R$};
      %phi
      \draw[dashed, very thick, chaptergreen] (axis cs:40,0) arc [x radius=25, y radius=25, start angle=0, end angle = 35];
      \node[anchor=south west, chaptergreen]  (ip) at (axis cs:38,0.07)  {$\phi_0$};
    \end{axis}
  \end{tikzpicture}
  \hfill
  \begin{tikzpicture}
    \definecolor{colorip}{rgb}{0.5,0.5,0.5};
    \begin{axis}[
      unit vector ratio*=1 1.4 1,
      xmin=-20,
      xmax=125,
      ymin=-20,
      ymax=100,
      grid=major,
      grid style={dashed, gray!30},
      axis y line=center,
      axis x line=middle,
      xlabel=$z$,ylabel=$y$,
      scale=0.7,
      ]
      %ip
      \fill[colorip] (axis cs:0,0) ellipse [x radius=2, y radius=2];
      % helix
      \addplot [very thick, domain=10:10+2.8*pi/3,samples=50]({(x-10)*38 + 10},{52*(sin(deg((x-10)-pi/3)) + sin(deg(pi/3))) - 10});
      \draw[thick, -, chapterlightblue] (axis cs:0.,-10) -- (axis cs:10,-10) node [pos=1.2, right] {$z_0$};
      \draw[thick, <-, chapterlightblue] (axis cs:9.,-10) -- (axis cs:17.,-10);
      \draw[thick, ->, chapterlightblue] (axis cs:-7.,-10) -- (axis cs:1.,-10);
    \end{axis}
  \end{tikzpicture}
  \hfill
  \begin{tikzpicture}
    \definecolor{colorip}{rgb}{0.5,0.5,0.5};
    \begin{axis}[
      unit vector ratio*=1.5 1.4 1.,
      xmin=-20,
      xmax=125,
      ymin=-30,
      ymax=150,
      grid=major,
      grid style={dashed, gray!30},
      axis y line=center,
      axis x line=middle,
      xlabel=$z$,ylabel=$s$,
      scale=0.7,
      ]
      %ip
      \fill[colorip] (axis cs:0,0) ellipse [x radius=2, y radius=2];
      \draw[very thick, -, black] (axis cs:10,0) -- (axis cs:120,140);
      \draw[dashed, thick, black] (axis cs:10,0) -- (axis cs:10,100);
      %lambda
      \draw[dashed, very thick, chapterorange] (axis cs:61,66) arc [x radius=84, y radius=100, start angle=53, end angle = 90];
      \node[anchor=south west, chapterorange] (ip) at (axis cs:29,77) {$\lambda$};
      \draw[thick, -, chapterorange] (axis cs:61,0) -- (axis cs:61,65) node [pos=0.5, right] {$\Delta s$};
      \draw[thick, -, chapterorange] (axis cs:10,0) -- (axis cs:61,0) node [pos=0.65, above] {$\Delta z$};
      % \draw[dashed, thick, chapterorange] (axis cs:61,41) arc [x radius=25, y radius=25, start angle=-90, end angle = -130];
    \end{axis}
  \end{tikzpicture}
  \caption{
    A schematic representation of the track's trajectory in the $x$\nobreakdash-$y$ (left), $z$\nobreakdash-$y$ (middle) and $z$\nobreakdash-$s$ (right) projections.
    All dimensions are in \SI{}{\cm}.
    The track parameters are: $d_0$, the signed distance of the closest approach to the $z$ axis (POCA);
    $\phi_0$, the angle defined by the $x$ axis and the track transverse momentum at the POCA; $z_0$, the $z$~coordinate at the POCA; and $\lambda$, the track dip angle.
    Also shown is the track radius $R$, which is the inverse of the absolute value of the track curvature $\omega$.
  }
  \label{fig:track_parametrization}
\end{figure}
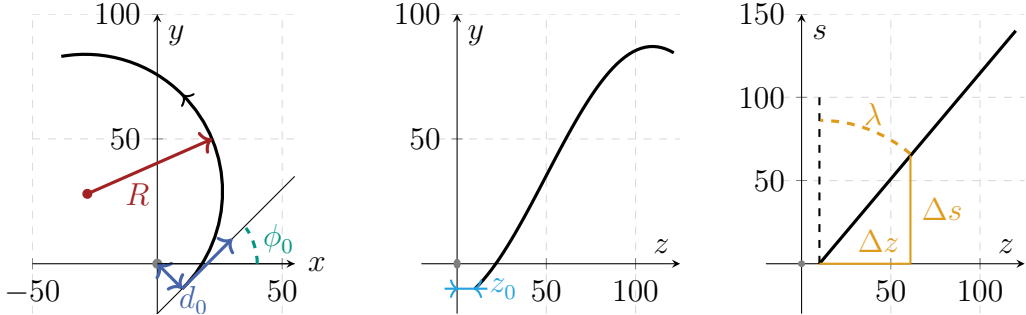

The \BelleII software uses data processing \emph{modules}, written in C++, which are loosely coupled and transfer data via a common exchange container.
This allows for the reconstruction task to be split into different sub-tasks which can be placed into a chain of independent and interchangeable modules performing the corresponding task.
The \texttt{RecoTrack} class is used as a common exchange format between algorithms to transfer track candidates from the different tracking detectors and their respective hits or clusters.
The final output of the track reconstruction is the \texttt{Track} class, which provides the fitted track parameters for the analysis user.

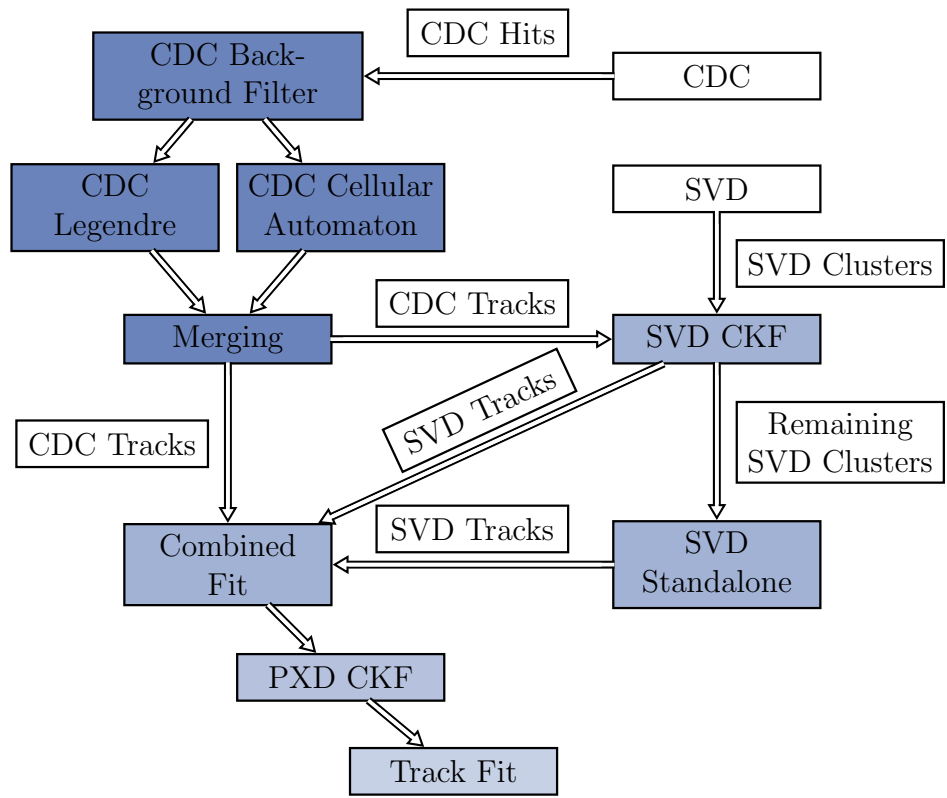
\begin{figure}
  \centering
  \begin{tikzpicture}
    \node[pathelement, fill=chapterblue!80!white, text width=8em] at (1.5, 2.0) (background) {CDC Background Filter};
    \node[pathelement, fill=chapterblue!80!white, text width=6em] at (0.0, 0.25) (legendre) {CDC Legendre};
    \node[pathelement, fill=chapterblue!80!white, text width=6em] at (3.0, 0.25) (ca) {CDC Cellular Automaton\vphantom{g}};
    \node[pathelement, fill=chapterblue!80!white, text width=6em] at (1.5, -1.5) (merger) {Merging};
    \node[pathelement, fill=chapterblue!50!white, text width=6em] at (8.0, -1.5) (svd_ckf) {SVD CKF\vphantom{g}};
    \node[pathelement, fill=chapterblue!50!white, text width=6em] at (1.5, -4.5) (ckf_merger) {Combined Fit};
    \node[pathelement, fill=chapterblue!50!white, text width=6em] at (8.0, -4.5) (svd_vxdtf2) {SVD Standalone\vphantom{g}};
    \node[pathelement, fill=chapterblue!40!white, text width=6em] at (3.0, -6.0) (pxd) {PXD CKF};
    \node[pathelement, fill=chapterblue!30!white, text width=6em] at (4.5, -7.25) (fit) {Track Fit};
    % \node[state, draw=none] at (8.0, 0.5) (svd_clusters) {};
    % \node[state, draw=none] at (8.0, 2.0) (cdc_hits) {};
    \node[pathelement, fill=white, text width=6em] at (8.0, 0.5) (svd_clusters) {SVD};
    \node[pathelement, fill=white, text width=6em] at (8.0, 2.0) (cdc_hits) {CDC};

    \draw[vecArrow] (cdc_hits) -- node[pathelement, above=0.25, midway, text width=4.5em] {CDC Hits} (background);
    \draw[vecArrow] (background) -- (legendre);
    \draw[vecArrow] (background) -- (ca);
    \draw[vecArrow] (legendre) -- (merger);
    \draw[vecArrow] (ca) -- (merger);
    \draw[vecArrow] (svd_clusters) -- node[pathelement, right=0.25, midway, text width=6.em] {SVD Clusters} (svd_ckf);
    \draw[vecArrow] (merger) -- node[pathelement, above=0.15, midway, text width=5.5em] {CDC Tracks} (svd_ckf);
    \draw[vecArrow] (svd_ckf) -- node[pathelement, right=0.25, midway, text width=6.em] {Remaining SVD Clusters} (svd_vxdtf2);
    \draw[vecArrow] (svd_ckf) -- node[pathelement, above=0.15, rotate=+25, midway, text width=5.5em] {SVD Tracks} (ckf_merger);
    \draw[vecArrow] (svd_vxdtf2) -- node[pathelement, above=0.15, midway, text width=5.5em] {SVD Tracks} (ckf_merger);
    \draw[vecArrow] (merger) -- node[pathelement, left=0.25, midway, text width=5.5em] {CDC Tracks} (ckf_merger);
    \draw[vecArrow] (ckf_merger) -- (pxd);
    \draw[vecArrow] (pxd) -- (fit);
  \end{tikzpicture}
  \caption{
    Overview of the steps performed for track reconstruction at \BelleII. See text for more details.
  }
  \label{fig:overview:schema}
\end{figure}

The track trajectories are represented locally using the helix parameterization, see \cref{fig:track_parametrization}.
The three helix parameters in the $x$\nobreakdash-$y$ plane are: the signed distance of the point of closest approach (POCA) to the $z$ axis, $d_0$;
the angle defined by the $x$ axis and the track transverse momentum at the POCA
, $\phi_0$; and the track curvature signed with the particle charge, $\omega$.
The helix can be represented by a straight line in the $s$\nobreakdash-$z$ space, with $s$ being the path length along the circular trajectory in the $x$\nobreakdash-$y$~projection.
The two corresponding parameters are: the $z$ coordinate at $d_0$, $z_0$; and the tangent of the dip angle $\tan\lambda$.

\cref{fig:overview:schema} shows an overview of the steps performed for track reconstruction at \BelleII.
Due to the very different properties of the three tracking detectors, different algorithms are used for each of them.
As a first step, the measured signals in the CDC are filtered and reconstructed by two independent algorithms:
a global track finding based on the Legendre~\cite{Alexopoulos2008} algorithm and a local algorithm employing a cellular automaton.
The results of both algorithms are merged and the CDC-only tracks are fitted employing a deterministic annealing filter (DAF)~\cite{Bilka2019}.
A combinatorial Kalman filter (CKF) is used to enrich the CDC tracks with SVD clusters.
High-curvature tracks that did not produce enough hits in the CDC are reconstructed with a standalone SVD track finder using an advanced filter concept called \SectorMap{} and a cellular automaton.
The results are combined, fitted again with a DAF and extrapolated to the PXD with a second CKF.
At this step the track finding stage is complete.
The following sections describe these steps in more detail.

The final step after the track finding includes a track fit using the DAF provided by the \genfit{}~\cite{Bilka2019} package.
For the fit, a specific particle hypothesis must be assumed to calculate the energy loss and the material effects correctly.
In \BelleII, all reconstructed tracks are fitted with the $\Ppi$, $\PK$ and $\Pp$ hypotheses.
The results of the fit is stored to be used in physics analyses.

% !TeX encoding = UTF-8
% !TeX spellcheck = en_US
% !TeX root = ./TrackingPaper.tex

\section{CDC Algorithm} \label{sec:cdc_algorithm}

Two distinct algorithms are used for the track finding in the CDC: global, and local track finding.
This enables a high track-finding efficiency while keeping the fake rate low.
The global track finding searches for patterns of hits consistent with helix trajectories, even with missing hits,
while the local track finding detects extended patterns of nearby hits.

Both algorithms make use  of the specific geometry of the CDC and exploit the flexibility of the software framework.
The software is written in a modular manner allowing for different sequences of algorithms.
Currently, the global track finding is performed first, after the initial filtering of the CDC hits.
Thus the global algorithm serves as the primary finding algorithm, which is followed by the local track finding algorithm.
The latter helps with reconstructing displaced tracks which originate far away from the interaction point.
The track candidates of both algorithms are then merged and post-processing is performed to remove falsely attached hits and, potentially, to attach additional ones.
The reconstructed tracks are then passed to the DAF algorithm to be fitted.

\subsection*{Global CDC Track Finding}

The global track finding in the CDC is based on the Legendre transformation~\cite{Alexopoulos2008}.
It is first performed in the $r$\nobreakdash-$\phi$~plane, using wire information from  axial layers only.
After that, it is extended to the three-dimensional space, by attaching wires from stereo layers to existing $r$\nobreakdash-$\phi$~trajectories.
The primary target of the algorithm is finding tracks originating from the vicinity of the origin in $r$\nobreakdash-$\phi$.
It is adjusted to identify also slightly offset tracks.

In the first step of the algorithm, the position information in axial layers is approximated by drift circles.
These drift circles are calculated using a calibrated $x$\nobreakdash-$t$~relation and time information corrected for particle time-of-flight and signal propagation time along the sense wire.
For the time propagation correction, it is assumed that particle trajectories are straight lines from the origin, that
particles travel with the speed of light, and that they cross the sense wires in the middle.
These assumptions are revised when the track parameters are determined.

The reconstruction in the $r$\nobreakdash-$\phi$~plane continues with a conformal mapping with the center at the origin.
This operation transforms circular trajectories starting from the origin to straight lines while the drift circles remain circles.
The track finding in the conformal space is thus reduced to the determination of straight lines tangential to a set of circles.

\begin{figure}
  \centering
  \subfloat[Two-dim.\ binary search]{
    \includegraphics[width=0.34\textwidth]{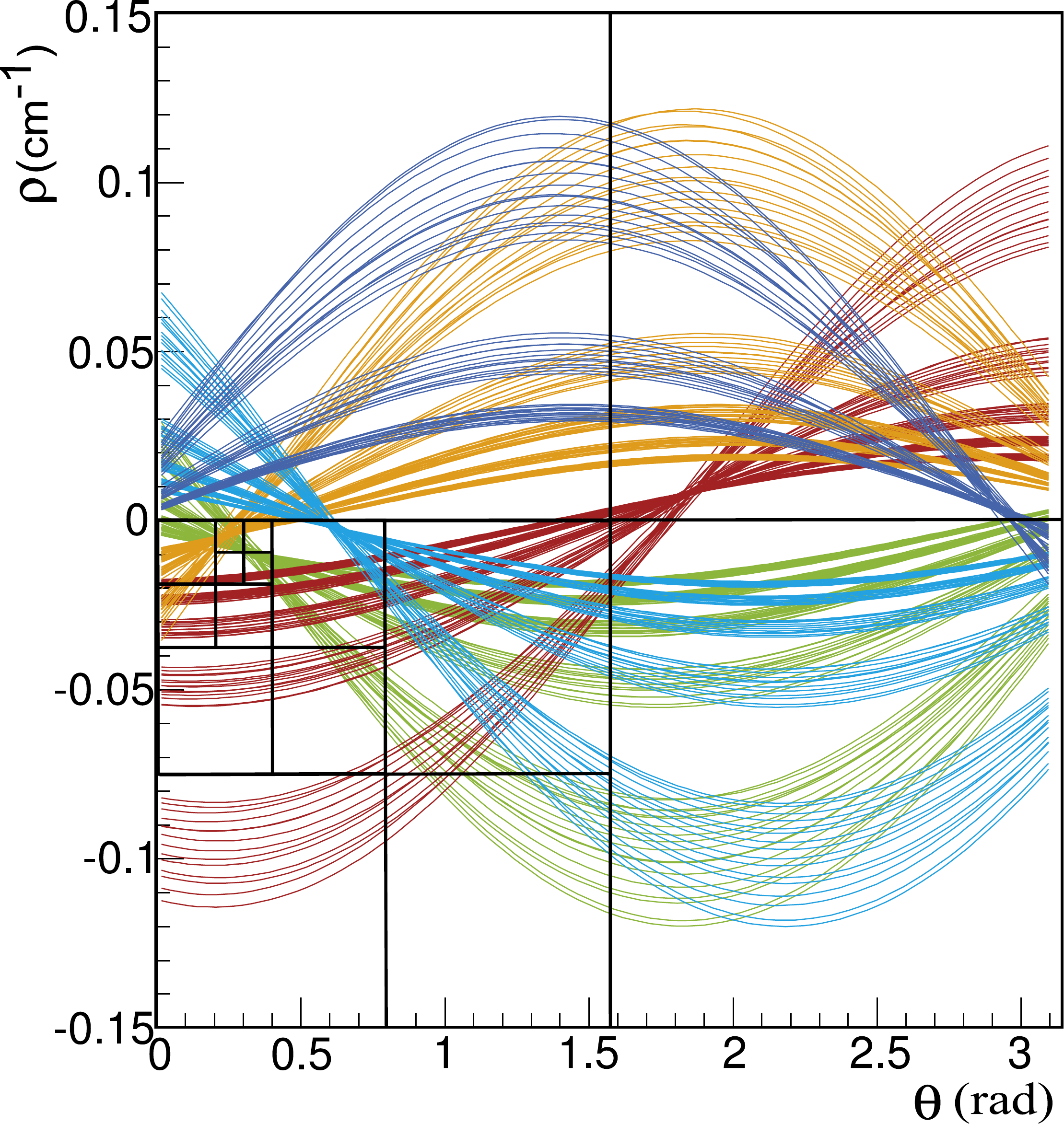}
    \label{fig:sliding_bins_a}%
  }
  \hfill%
  \subfloat[Sliding bins example]{
    \includegraphics[width=0.615\textwidth]{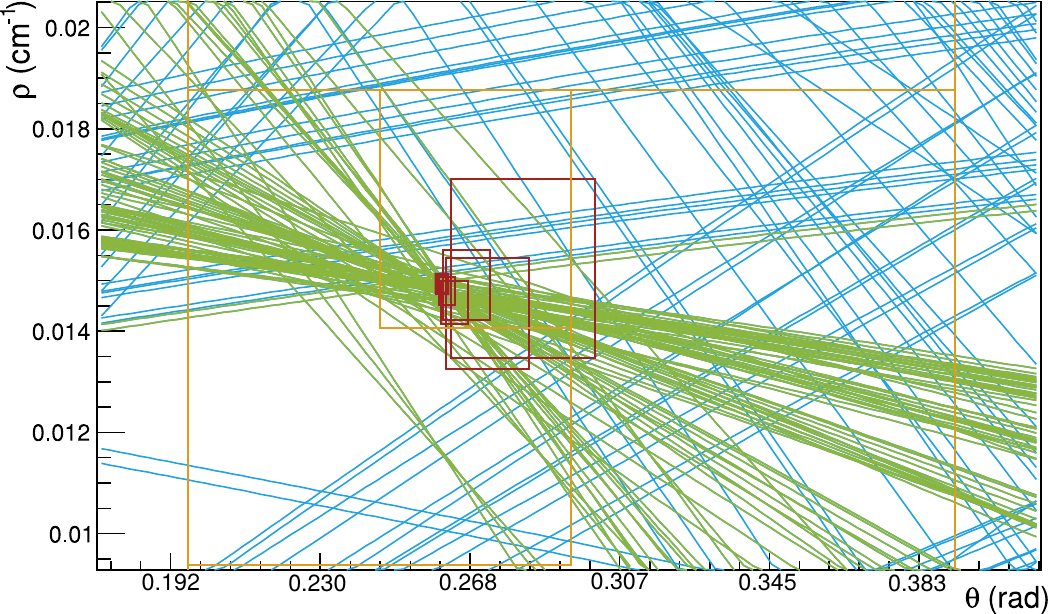}
    \label{fig:sliding_bins_b}%
  }
  \caption{
    Examples of:
    (a) standard two-dimensional binary search algorithm;
    (b) modified algorithm with variable bin size.
    See text for more details.
  }
  \label{fig:sliding_bins}
\end{figure}

The equation of a tangent to a drift circle in conformal space can be represented  using the two Legendre parameters $\rho$ and $\theta$ as
$$
  \rho = x_0 \cos \theta + y_0 \sin \theta \pm R_{\mathrm{dr}}\,,
$$
where $(x_0,\, y_0)$ and $R_{\mathrm{dr}}$ represent the center of the circle and its radius, respectively.
Hence, each drift circle maps to a pair of sinusoids in the $\rho$\nobreakdash-$\theta$ track-parameter space.
The track recognition and track parameter determination correspond to finding the most populated regions in the $\rho$\nobreakdash-$\theta$~space.
An efficient method to localize these regions is a two-dimensional binary search algorithm, as illustrated in \cref{fig:sliding_bins_a}.
The algorithm consists of dividing the $\rho$\nobreakdash-$\theta$~space into four equally sized bins and selecting the most populated of them for further subdivision, until convergence.

The two-dimensional binary search algorithm uses a dedicated \emph{quadtree} data structure~\cite{article:quadtree} to store intermediate search results.
Each node in the quadtree is linked to four children, corresponding to four sub-bins of the node.
In general, the search is continued only for the sub-bin containing the most hits.
However, it is possible to step back and examine other directions, without repeating the search from the beginning, which speeds up the search for multiple track candidates.

The binary search stops when the bin size becomes smaller than a resolution parameter that is taken to be dependent on $\rho$.
This accounts for the smearing of the track parameters due to the energy loss, non-uniformity of the magnetic field, displaced IP, uncertainty of the drift circle radii and wire displacements.
The resolution function is optimized using simulated events.

The introduction of the  resolution function as the stopping criterion allows to extend the algorithm to non-standard bin sizes.
For a track that is displaced from the origin, the crossing points in the Legendre space may be split between two bins.
This effect can be reduced greatly by allowing for overlapping bins.
Bins extended by \percentageNumber{25} with respect to the exact division are used.
A positive side effect of this feature is that the overlapping bins tend to \emph{slide} towards the maximal density of intersections, as illustrated in \cref{fig:sliding_bins_b}.

Multiple tracks are found iteratively, using several passes over the Legendre space.
At each pass a new track candidate is declared to be found when it satisfies certain quality criteria, such as the number of attached hits.
These quality criteria can be varied to increase finding efficiency for different track topologies.
Hits corresponding to the found track candidates are removed from further iterations.
The high-momentum tracks crossing all CDC layers are searched for first, followed by curling tracks and tracks with large longitudinal momentum, which leave the chamber at smaller radii.

The $r$\nobreakdash-$\phi$~track candidates are subjected to a post-processing step,
performed in the physical $r$\nobreakdash-$\phi$~space using the fast fitting algorithm of~\cite{Karimaki1991}.
Firstly, the track candidates are checked to see if they can be merged, to reduce the clone rate.
This includes the merging of individual loops from curling tracks.
The merge algorithm uses a $\chi^2$-based criterion, comparing the quality of the circular fits to the hits from the separate track-candidates to the fit to the combined set of hits.
In addition, hits from the track candidates are examined to determine if they have to be removed or re-assigned to other tracks.
Finally, all unassigned hits are checked to determine if they can be attached to the existing track candidates.

Hits from stereo layers, containing $z$ information, are added to the $r$-$\phi$ trajectories at the next step.
The $r$\nobreakdash-$\phi$~trajectory is used to reconstruct the position information of each stereo measurement.
As the stereo wire can be approximated by a straight line and the drift circle does not have direction information, finding the position gives two solutions:
either the drift circle is enclosed by the trajectory circle, or not --- giving two possible position values for each hit.
Stereo hits with a reconstructed $z$~coordinate $z_\text{rec}$ determined far outside the detector volume are dismissed.
Given that $z_\text{rec}$ depends strongly on the estimated $r$\nobreakdash-position of the trajectory at the stereo wire which may be not very accurate,
hits as far as twice the physical drift chamber length are retained. 

The problem of track finding becomes very similar to the search in the conformal mapping of the $r$\nobreakdash-$\phi$~space which makes it possible to use the same algorithm as described above.
This time, the trajectory is straight in the $s$\nobreakdash-$z$~space and 
it can be described by the equation
$$
  z_0 = z_\text{rec} - \tan \lambda \cdot s_\text{rec}\,,
$$
with $s_\text{rec}$ being the path to the stereo-wire hit.
This gives a line of possible trajectory parameters $(z_0, \tan \lambda)$ for each stereo-layer hit.
The point with the most intersections of the lines in the parameter space is used to determine the track parameters.
For this, an analogous implementation of the quadtree algorithm as described above is used.

The stereo-wire hits that are found are added to the $r$\nobreakdash-$\phi$~track only in the case they are not selected for another $r$\nobreakdash-$\phi$~track.
On average, 19\% of the hits are compatible with more than one of the tracks.
In these cases the hits are not added to any track which increases the purity of the hit assignment.

\subsection*{Local CDC Track Finding}
To complement the global search approach, and to detect short tracks and tracks displaced from the IP with a high efficiency,
the local track finder operates without any assumption on the origin of tracks.
The algorithm searches for connected hits in the CDC superlayers, so called segments.
This search uses the cellular automaton concept, which acts on an acyclic graph of vertices connected by edges.
More specifically, a weighted cellular method is used, where the vertex $i$ has the weight $\Theta_i$ and the edge between the vertices $i$ and $j$ has the weight $w_{ij}$.
Now, track finding can be formulated as maximization of an energy function which can be formulated with
$$
  E_i = \sum{w_{ij}} + \sum{\Theta_j} \,,
$$
where the sums are taken along a path to the vertex $i$.
The concept of the weighted cellular automaton is employed in two different stages:

\begin{figure}[t]
  \centering
    \includegraphics[width=0.7\textwidth]{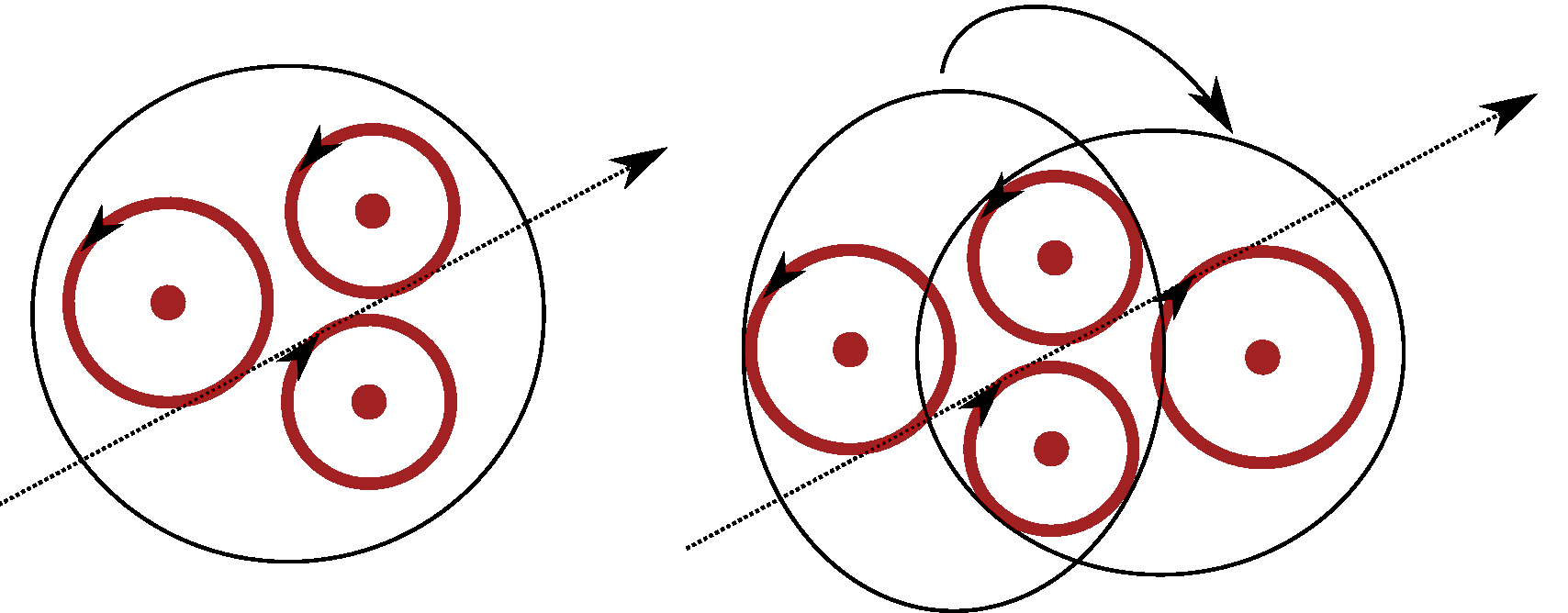}
  \caption{Combination of three neighboring wire hits to a graph vertex (left) and two triplets sharing two wire hits to a graph edge (right).}
  \label{fig:local_finder_segments}
\end{figure}

\paragraph{Segment building stage}
Vertices (\emph{triplets}) are formed by combining three neighboring hits  and assuming the left-right passage hypotheses
for a unique trajectory through these three hits\footnote{In general, several triples are built for a given set of three neighboring hits,
depending on  the left-right passage hypothesis.}.
A linear trajectory is then extracted from the measured drift circles by a least-squares method and the weight $\Theta_j$ is assigned based on the $\chi^2$ value of the fit.
Edges are created from neighboring triplets that share two hits and which pass loose feasibility cuts,
with the weight $w_{ij}$ determined based on the $\chi^2$ value of the straight line fit to the four drift circles (see \cref{fig:local_finder_segments}).
The algorithm allows for information missing from one CDC layer.

\begin{figure}[t]
  \centering
    \includegraphics[width=0.65\textwidth]{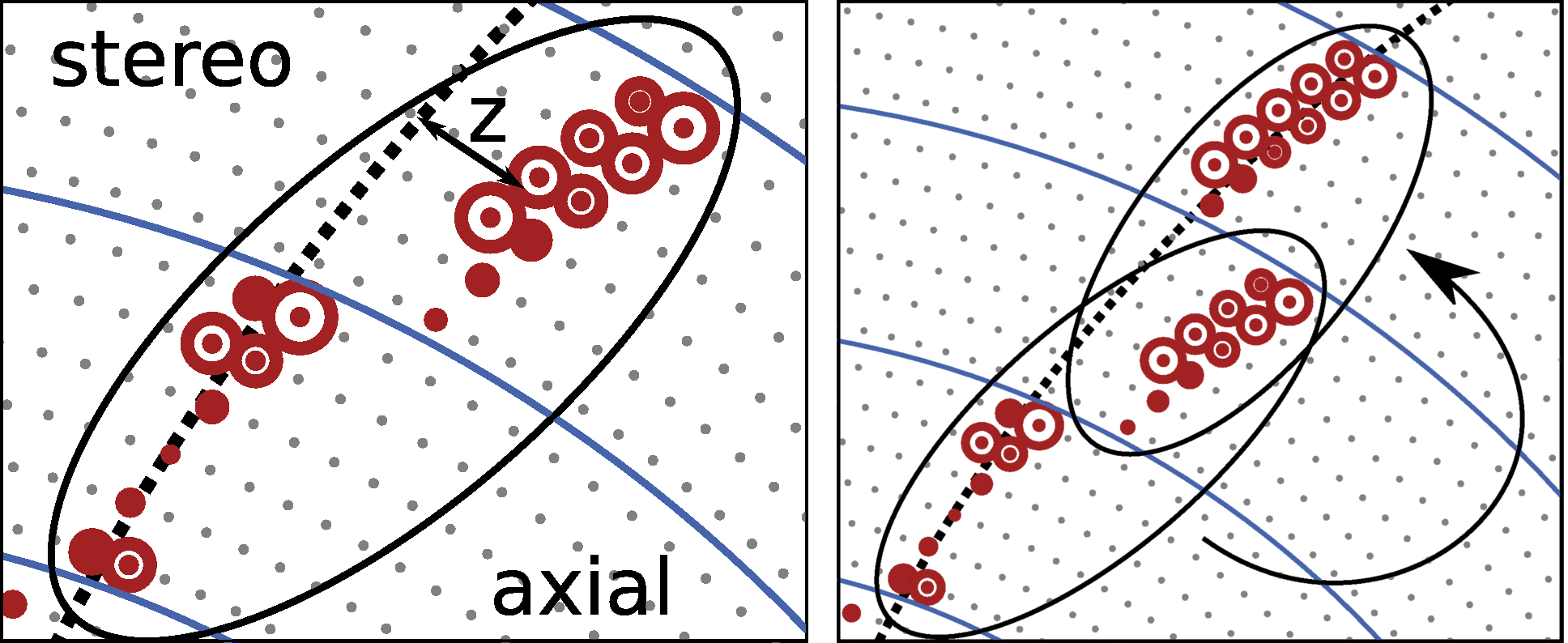}
  \caption{Combination of a pair of axial- and stereo-wire segments to one graph vertex (left) and the combination of vertices, that share one segment, to a graph edge (right).}
  \label{fig:local_finder_track_finding}
\end{figure}

\paragraph{Track building stage}
This stage combines the individual segments found in the axial and stereo superlayers to longer tracks.
The vertices are created from a pair of segments in neighboring axial- and stereo-wire superlayers.
The weight $\Theta_j$ of each vertex is computed with a $\chi^2$ circle fit using the Riemann method~\cite{Strandlie2000}
and the reconstruction of the $z$~coordinate is performed using a linear fit in the $s$\nobreakdash-$z$~space.
Neighboring vertices that share one segment form the edges in the cellular automaton's graph (see \cref{fig:local_finder_track_finding}).
The corresponding weight $w_{ij}$ is computed based on the $\chi^2$ value of the fit to hits from all segments.
Additional information, such as the number of hits per segment, can be included in the weight calculation using multivariate analysis methods.

\subsection*{Combination of Local and Global Tracking Results}
The two track finding algorithms described above are both used to find tracks from the full set of CDC hits.
This is done to exploit their specific benefits, with the global track finding  capable to reconstruct tracks with several missing layers
and the local track finding having similar efficiency regardless of the track origin.

For combining the results of both tracking approaches, the track candidates from the global track finder are used as a baseline.
Segments found by the local track finder are added to those tracks using a multivariate approach.
The track-segment combination is based on FastBDT~\cite{keck2016}, which uses several variables calculated from the track and the segment
(e.g.\ the number of common hits, helix parameters, hit-to-trajectory distances) into one single number, which classifies between correct and wrong matches.
The multivariate method is trained using simulated events.

Several quality filters based on multivariate estimators are applied to the found tracks and their hits.
This increases the hit purity, improves the track parameter resolution, and decreases the rate of  fake and clone tracks.

% !TeX encoding = UTF-8
% !TeX spellcheck = en_US
% !TeX root = ./TrackingPaper.tex

\section{SVD Standalone Algorithm} \label{sec:vxd_algorithm}
 % General Introduction
 A dedicated standalone algorithm is employed for the task of track finding with the SVD.
 This algorithm reconstructs the low momentum particles with a transverse momentum of less than \SI[per-mode=symbol]{100}{\mega\eV\per c} which deposit too few hits in the CDC.
 However, due to the proximity of the SVD to the beam, the algorithm has to cope with a high occupancy from beam-induced background.
 The original idea and implementation for this algorithm, called the VXD Track Finder (VXDTF), is described in~\cite{Lettenbichler2016}.
 Further improvements of the algorithm, which are partly described in~\cite{Wagner2017} and~\cite{Racs2018}, led to its second version, VXDTF2.

 % The input to the algorithm (space points) and a overview of the algorithm
 The input to the VXDTF2 algorithm is the set of the three dimensional \SpacePoints{} created in the pre-processing steps described in \cref{sec:input} from the SVD measurements\footnote{
 It is also possible to use the three dimensional measurements provided by the PXD.
 However, due to the combined effect of its proximity to the beam line and the lower readout frequency of its sensors, the PXD is subject to a substantially higher occupancy from beam-background induced hits.
 Omitting the PXD measurements from the track-finding process simplifies the combinatorial problem and leads to a purer set of track candidates produced by the standalone algorithm.
 The task of evaluating the additional information available in the measurements of the PXD is passed on to a dedicated algorithm described in \cref{sec:ckf_description}.
 }.
 The VXDTF2 algorithm consists of three steps.
 In the first step, graphs of related \SpacePoints{} are created using geometrical information.
 A map, called a \SectorMap{}~\cite{Fruhwirth2013}, containing the geometrical relations between different regions of the silicon detector as well as additional selection criteria, supplies the necessary input for this step.
 The prepared graphs are then evaluated in the second step by a cellular automaton which yields a set of paths.
 As third step, the final set of SVD track candidates is chosen by selecting the best paths.

\subsection*{\SectorMap{}}
\begin{figure}
  \centering
  \includegraphics[width=0.5\textwidth]{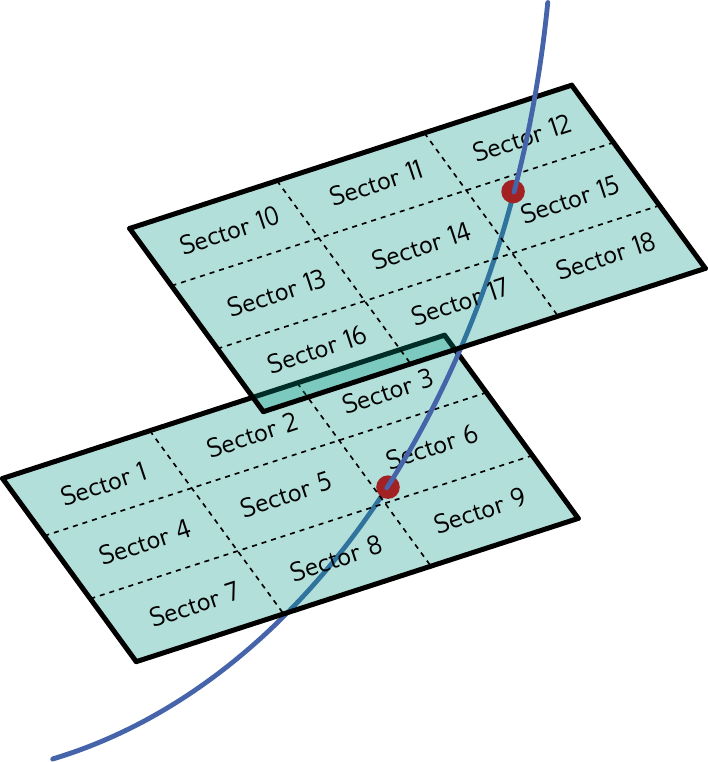}
  \caption{Illustration of the subdivision of two sensors into nine sectors each and of the relation between the sectors 6 and 15 which are traversed by the same track.}
  \label{fig:SectorMapSectors}
 \end{figure}

 % Sectors on Sensors Concept
 The \SectorMap{} is a data structure that holds information about how \SpacePoints{} in different regions of the detector can be related by tracks.
 To cope with the high number of possible combinations of \SpacePoints{}, the sectors on sensor concept --- originally proposed in~\cite{Fruhwirth2013} --- is used for track finding with the SVD.
 This concept consists in subdividing each sensor into smaller sections, called \emph{sectors}.
 The default setup is a division of each sensor element into three parts along its width and three parts along its length, resulting in nine sectors per sensor.
 It is possible to adjust the number of sectors individually for each sensor, which allows the granularity to be adapted to changing detector conditions during the run time of \BelleII.
 This representation of the SVD geometry allows to define directed relations between sectors of the detector which commonly contain measurements of the same track, as illustrated in \cref{fig:SectorMapSectors}, where the sectors 6 and 15 are related due to the track traversing both sectors.
 The direction of the relations is defined by the order in which the sectors are traversed.
 A mapping among sectors defined by these relations allows for a significantly reduced number of combinations of \SpacePoints{} as input to the algorithm.
 
 % Filters
 In addition to the relations between sectors of the detector geometry, the \SectorMap{} holds selection criteria to be fulfilled by combinations of \SpacePoints{} on related sectors.
 These criteria are called \emph{filters} and are defined for pairs of two as well as triplets of three \SpacePoints{}.
 They provide a way to reject background hits and thereby a further reduction of the \SpacePoint{} combinations to be evaluated per event.
 Each filter is a function which calculates a specific quantity, called \emph{filter variable}, for a given \SpacePoint{} pair or triplet, and checks if the result is within a given validity range.
 The validity range depends on the filter variable and on the sectors.
 It is stored for each individual sector combination alongside the respective relation between sectors in the \SectorMap{}.
 Filter variables are mostly geometrical quantities derived from the spatial information of the \SpacePoints{}, or are calculated from the precise timing information provided by the SVD.

\begin{figure}
  \centering
  \includegraphics[width=0.94\textwidth]{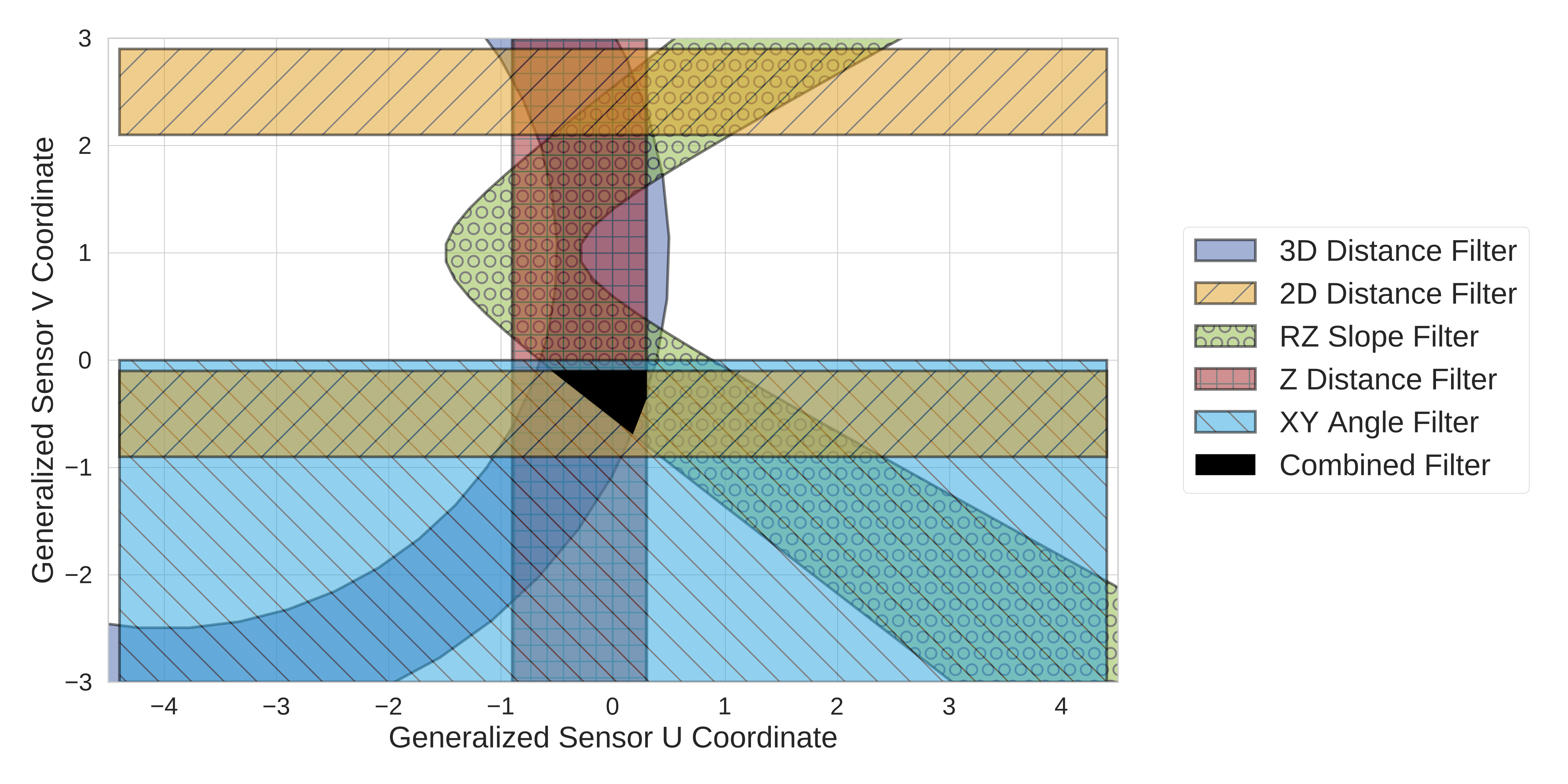}
  \caption{
    Illustration of the selection power of the combination of several \SectorMap{} filters.
    The illustration depicts a sensor plane and the areas selected by several \SectorMap{} filters calculated for a given \SpacePoint{} on another sensor layer.
    Only \SpacePoints{} within these areas are allowed to be combined with the given \SpacePoint{} on the other layer.
    The combination of all filters reduces the allowed region to the area shown in black.
  }
  \label{fig:SectorMapCuts}
 \end{figure}

 % 2-SP Filters
 The variables calculated for filters for \SpacePoint{} pairs are simple quantities
 such as distances between the two \SpacePoints{} (in one-, two-, and three-dimensions),
 angles in $\phi$- and $\theta$-direction defined by the two \SpacePoints{},
 or the difference in their detection times.
 An illustration of the combined application of a selection of such filters is depicted in \cref{fig:SectorMapCuts}.

 % 3-SP Filters
 More complicated quantities can be evaluated for filters based on the combination of three \SpacePoints{}.
 These include for example the angle enclosed by the two segments defined by the three \SpacePoints{}, 
 or the position of the center as well as the radius of the circle defined by the three \SpacePoints{} in the  $x$\nobreakdash-$y$~plane.
 \SpacePointCapital{} triplet filters based on the SVD timing information are also employed.
 As the SVD is composed of only four layers and a triplet of \SpacePoints{} already provides enough degrees of freedom to unambiguously define a helix trajectory of a charged particle in a magnetic field, further filters for combinations of four and more \SpacePoints{} are not considered.

 % Training Process
 The directed relations between sectors as well as the filter selection criteria are obtained via a training process based on Monte Carlo events.
 For this purpose, a dedicated sample of representative \PUpsilonFourS{}~events is generated.
 Track candidates are selected from this training sample using the MC track finder (see \cref{sec:simulation}).
 Additional samples of high-momentum tracks such as simulated high-momentum muon events or simulated Bhabha events can be incorporated into the training process as \PUpsilonFourS~events don't typically produce such tracks.
 Based on this set of tracks, directed relations between pairs of sectors are obtained for all pairs of sectors which have been traversed subsequently by at least one track.
 The selection ranges for the filters are defined by the minima and the maxima or by quantiles of the distributions of the respective filter variables as observed during the training for each individual sector combination.

 This training process allows the \SectorMap{} to learn the geometry of the SVD setup.
 Hence, it can adapt to changing detector conditions like defects on the sensors or even the loss of complete sensors or layers, as long as these defects are modeled by the simulation.
 The \SectorMaps{} produced in this manner are stored in the database of \BelleII, which allows defining different \SectorMaps{} for different run conditions.

\subsection*{Track Finding Algorithm}
 To address the high combinatorics during the process of building track candidates from the \SpacePoints{} provided by the SVD, the \SectorMap{} filters are used.
 A first directed graph, called \emph{sector network}, is build for an event with all active sectors (sectors on which hits are detected) as nodes.
 The edges connecting the nodes of the sector network are given by the directed relations stored for the respective active sectors in the \SectorMap{}.

 The next step comprises the creation of a second directed graph with the \SpacePoints{} on the active sectors of an event as nodes.
 The edges of this \emph{\SpacePoint{} network} are given by the edges of the sector network and connect \SpacePoints{} in pairs if they pass the criteria of the \SectorMap{} filters for \SpacePoint{} pairs.
 Their directions are defined by the respective edges in the sector network.
 The resulting \SpacePoint{} pairs are called \emph{segments}.
 Next, pairs of such segments that have a \SpacePoint{} in common are combined into triplets of \SpacePoints{}, creating a third directed graph with segments as nodes, which is therefore referred to as a \emph{segment network}.
 Again, the criteria given by the respective filters for \SpacePoint{} triplets provided by the \SectorMap{} are considered during the combination of the segments to triplets.

 All paths given by the edges of the segment network with a minimal length of three \SpacePoints{} are considered as track candidates without further restrictions from the \SectorMap{}.
 A cellular automaton is used to gather the longest paths in the graph, beginning with the nodes on the outermost layers as these are least occupied by beam-background induced hits.

 Gathered this way, the track candidates in an event may share SVD clusters or even \SpacePoints{}.
 At this stage of the algorithm, the number of fake and clone track candidates make up more than half of the track candidates and are directly related to the overlaps among a set of tracks.
 To reduce the fraction of fakes and clones, the final set of track candidates for an event is required to be composed only of candidates which do not share any SVD clusters among them.
 As roughly \percentageNumber{5} of all tracks in a normal \PUpsilonFourS{}~event share at least one SVD cluster with another track,
 this introduces a small loss of less than \percentageNumber{1} in finding efficiency to the benefit of an increase in purity for the final set of track candidates by a factor of roughly two.
 Any cases where two or more track candidates share common hits are resolved based on a rating of all track candidates, followed by a greedy local selection as explained in the following paragraphs.

 For the rating of each track candidate, a quality indicator determined from the goodness of a fast fit to the candidate is employed.
 The fit method used for this objective is an adapted version of the Triplet Fit introduced in~\cite{Berger2017}.
 This method is chosen because it takes into account the multiple scattering relevant for the tracks left by low momentum particles of interest to the VXD standalone track finding.
 The Triplet Fit is applied to each path supplied by the cellular automaton, as well as their subpaths obtained by excluding one or multiple \SpacePoints{} from the original path.
 The latter allows for the exclusion of misattributed \SpacePoints{} and results in a track candidate with higher purity.
 Furthermore, the inclusion of the subpaths can lead to a recovery of the efficiency loss due to overlapping true tracks.
 When creating the subpaths, the rule of a minimal length of three \SpacePoints{} is still obeyed.

 The Triplet Fit yields a $\chi^2$ value for each track candidate by combining fits to all \SpacePoint{} triplets contained within a candidate under consideration of the effect of multiple scattering.
 For this estimate the average radiation length of the SVD sensor material as reported in~\cite{Abe2010}, as well as a first approximation of the entrance angle of the particle with respect to the sensor plane are taken into account.
 The $p$-value is calculated for each track candidate from its $\chi^2$ value and degrees of freedom and used as a quality indicator.
 Based on these quality indicators, the final set of non-overlapping track candidates is chosen via a greedy selection which takes the candidate with the highest quality indicator among the ones competing for a \SpacePoint{}.

 Optionally, a multivariate method can be applied which combines the acquired quality indicator with further features,
 such as a particle momentum estimate, the number of \SpacePoints{} and properties of the involved SVD clusters.
 This approach can yield a quality indicator with an enhanced performance.
 For this purpose a FastBDT is trained on Monte Carlo events obtained by applying the candidate creation steps up to the point of the Triplet Fit.
 In the resulting training sample, the track candidates with a purity of \percentageNumber{100} are marked as truth target.
 Therefore, a FastBDT trained in this manner has learned to identify track candidates with a high purity.
 Enabling this auxiliary multivariate analysis method for the overlap removal increases the achieved track finding efficiency, albeit with a significant drop in hit efficiency.
 This option is not therefore used as a basis for the resolution of overlaps, but used to produce a refined track quality estimate in an additional step.
 This indicator of the track quality is stored for all tracks of the final set and can later be accessed and used in the event selection of physics analyses.

 The algorithm is further optimized as it is found to perform more slowly than acceptable at the HLT for certain peculiar Bhabha events.
 The slowdown is understood as follows.
 In rare cases, highly energetic electrons scatter in the material of the final focusing magnets, thereby causing a shower of secondary particles which leave a large number of clusters in a small area of the SVD.
 This leads to a significant increase in the combinatorics during the candidate creation process that cannot be restricted by the \SectorMap{} filters.
 To tackle this issue, two additional measures are incorporated into the candidate creation.
 Firstly, a limit on the number of nodes and edges in the three networks is introduced, as the problematic Bhabha events can mostly be identified based on noticeably high values for these quantities.
 If the limits determined on Monte Carlo simulations are exceeded, the processing of the event is aborted and the problematic event is marked as such.
 The limits are chosen so that the desired measurements of the \PUpsilonFourS{}~resonance are not affected.
 Secondly, during the path-collection step an additional selection procedure based on the segment network is applied, which evaluates overlaps already in this graph.
 All paths associated with a given segment are grouped together and evaluated with the Triplet Fit.
 Only a fixed number of best candidates from each group is considered for further processing steps.
 This early candidate reduction based on \SpacePoint{} pair overlaps imposes an additional limit on the combinatorics for problematic events which slip through the aforementioned limits.
 By means of these additional selection steps, the problematic events can be handled by the VXDTF2 and the run time limits imposed by the requirements of the HLT are satisfied.

% !TeX encoding = UTF-8
% !TeX spellcheck = en_US
% !TeX root = ./TrackingPaper.tex

\section{Combinatorial Kalman Filter} \label{sec:ckf_description}
% Introduction into CKF -> Kalman filter
The combinatorial Kalman filter (CKF) is widely used in tracking in high-energy physics experiments~\cite{Mankel1999, Mankel2004, CMSCollaboration2014, Aaboud}.
One of the advantages of the method is that it produces tracks with high purity also in environments of high hit densities.
The CKF is an iterative local algorithm and was first described in~\cite{Billoir1989}.
Starting with a seed estimation of the track parameters with uncertainties, the track is extrapolated with the Runge-Kutta-Nystr\"om method~\cite{Lund2009} into the detector volume.
Hereby, non-uniform magnetic fields are included in the numerical solution of the equation of motion.
A correction of the energy and the uncertainties due to energy loss and multiple scattering can optionally be included.
After the extrapolation, possible hit candidates are determined based on the current position and uncertainties of the track candidate.
The next hit candidate is added to the track and the procedure is repeated.
If there are multiple mutually exclusive next-hit candidates, the whole track candidate is duplicated and subsequently treated as two tracks.
In the end, the final track candidate is selected according to different quality criteria.

% SVD CKF
As a first step, the track candidates found by the CDC track finding algorithm are used as seeds to attach SVD \SpacePoints{}.
Hereby, low momentum tracks can have both start and end points in the inner layers of the CDC, so both points can be used as a possible seed to account for wrongly assigned charges in the CDC.
The CDC seeds are fitted using a DAF algorithm assuming a pion mass hypothesis.
These seeds are iteratively extrapolated to the SVD sensors and SVD \SpacePoints{} are attached.
Material effects are disregarded at this stage to increase the processing speed.
Due to the complex detector layout, the different use cases, and the complex input data from the CDC track finding algorithm, the filter decisions in the CKF are taken by a FastBDT trained on simulated events including the beam-induced background.
Variables such as the distance between extrapolated and measured hit position as well as the calculated $\chi^2$ of the hit are taken into account.
They are enriched with information about the track candidate, for example the number of attached \SpacePoints{} or the estimated transverse momentum.
The number of track candidates that is considered is restricted to ten to keep the computational effort on a manageable level.
After a final candidate selection based on FastBDT using full-track information such as the summed $\chi^2$ and the number of missing layers, the combined CDC-SVD track is refitted using another DAF with a full material effect handling.

% Merging SVD CKF
Due to hit inefficiencies of the CDC algorithm, especially for the stereo layers, the track resolution can be extremely poor when $|z_0|$ is above \SI{1}{\centi\meter}.
Therefore, it is not possible to attach SVD \SpacePoints{} reliably to every reconstructed CDC track.
To solve this issue and to find additional low-momentum tracks, the VXDTF2 described in \cref{sec:vxd_algorithm} is applied to the set of remaining \SpacePoints{} in the SVD.
The merging of additional SVD candidates with unmerged CDC tracks is performed by a second pass of the CKF.
These unpaired CDC tracks are used as seeds and only \SpacePoints{} found by the VXDTF2 are allowed as input.
Compared to the first pass of the CKF with all \SpacePoints{}, simpler filters are applied during the processing due to the high purity of the VXDTF2 algorithm.

% PXD CKF
All reconstructed CDC-SVD tracks are then used to extract regions of interest in the PXD during the online reconstruction.
In the offline reconstruction, the PXD clusters collected in these regions of interest are used as input to the last application of the CKF and are attached to their combined CDC-SVD tracks.
The implementation is based on the same general principles as the SVD CKF.
It uses the same FastBDT filters which are now applied to the PXD clusters.
An additional input in the BDT classification is given by the position and the shape of the PXD clusters.

% Implications on the tracking results, processing time
The precise positions of the PXD clusters in the tracks improve the resolution on the spatial track parameters $d_0$ and $z_0$ by a factor of two and more.
The efficiency of attaching SVD (PXD) hits is over \percentageNumber{85} (\percentageNumber{89}).
The purity of the attached SVD or PXD hits is above \percentageNumber{98} and \percentageNumber{96}, respectively.

% !TeX encoding = UTF-8
% !TeX spellcheck = en_US
% !TeX root = ./TrackingPaper.tex

\section{Performance of the Track Finding} \label{sec:track_finding_performance}

The following section describes the performance of the tracking algorithms presented in this paper.
The performance is evaluated on an independent set of simulated \PUpsilonFourS{}~events, including beam-induced background simulated for the anticipated full instantaneous luminosity of \SI{8e35}{\per\centi\meter\squared\per\second}.
In the simulation, a detector setup with nominal positions is used.
The results of the reconstruction are compared to the respective MC tracks.
Quantities such as the purity or efficiency are calculated on this sample using the definitions from \cref{sec:simulation}.
All quoted uncertainties on these quantities are calculated using the method of bootstrapping~\cite{efron1979}.

\begin{figure}
  \centering
  \subfloat[On all particles.\label{fig:performance:fe:all}]{
    \centering
    \includegraphics[scale=0.33]{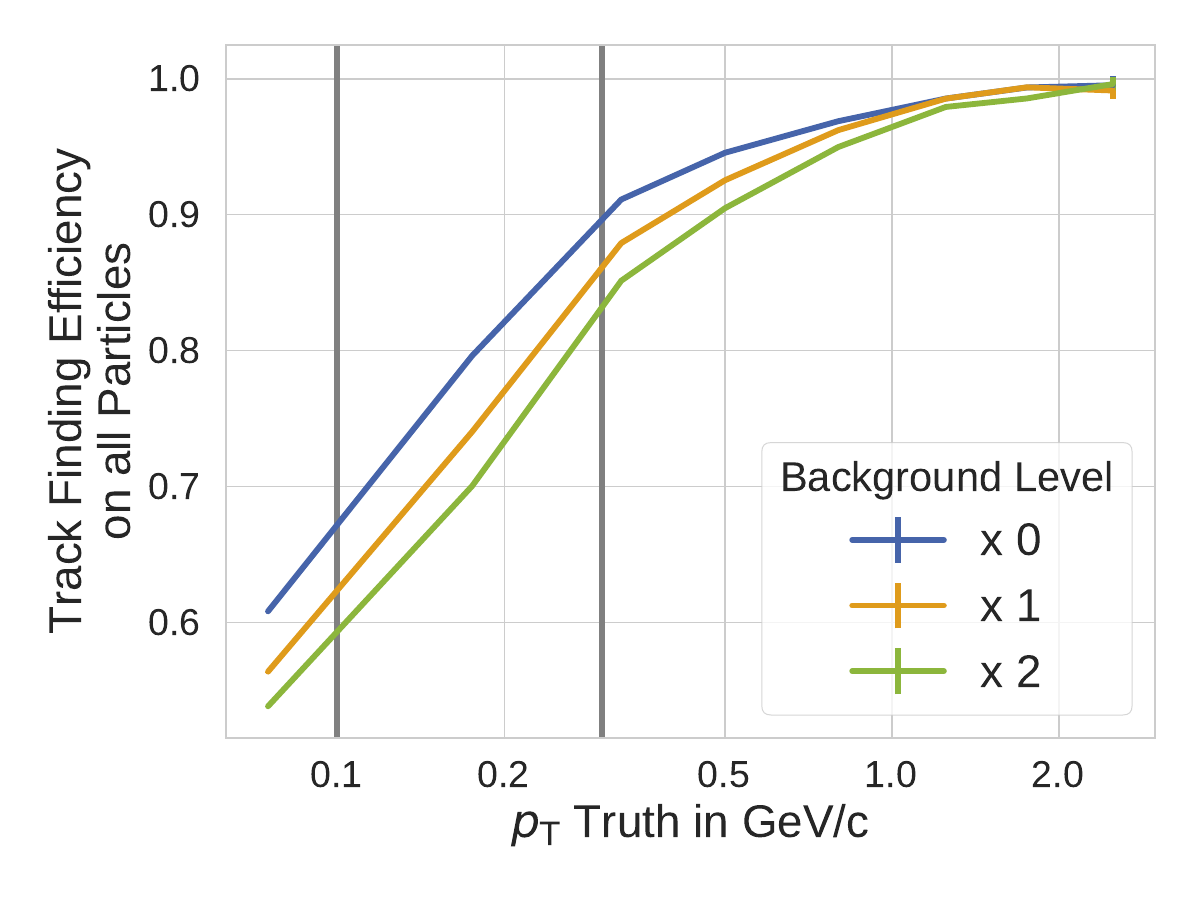}
  }
  \subfloat[Only on primaries.\label{fig:performance:fe:prim}]{
    \centering
    \includegraphics[scale=0.33]{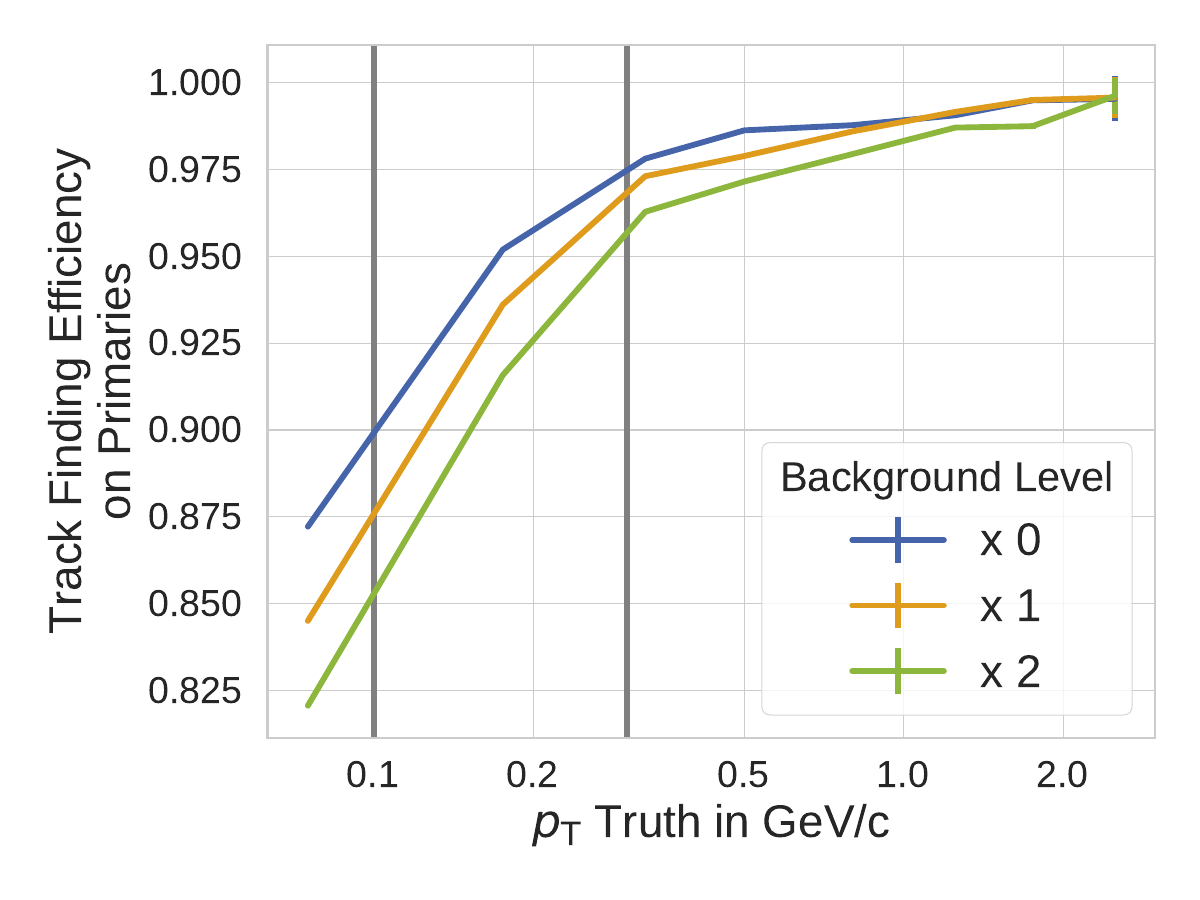}
  }
  \caption{
    Track finding efficiency calculated for simulated \PUpsilonFourS{}~events with different levels of beam-induced background relative to the expected level.
    \cref{fig:performance:fe:all} is calculated on all trackable simulated particles, whereas \cref{fig:performance:fe:prim} only takes into account trackable particles from the primary interaction.
    The gray vertical lines indicate the typical transverse momentum of particles only trackable in the VXD (below left line) and with high efficiency in the CDC (above right line).
  }
  \label{fig:performance:fe}
\end{figure}

\cref{fig:performance:fe} shows the track-finding efficiency for different simulated transverse momenta and different levels of beam-induced background relative to the anticipated level.
A distinction between final-state particles stemming from the primary $\Pep\Pem$ interaction and decays of short-lived particles, produced by event generators (\emph{primaries}),
and all final-state particles including those produced by Geant4 during the travel through the detector (\emph{secondaries}) is made.
Most of the analyses rely only on the former, whereas the latter can give valuable additional information for decays in flight or for particle identification.
For the \HepProcess{\PKs \to \Ppiplus \Ppiminus} decay, the efficiency depends strongly on the distance of the decay vertex from the IP.
For smaller distances it remains high, consistent with a product of efficiencies of the two primary tracks, for larger distances it drops below \percentageNumber{80}.
For \PKs{} particles produced in \PB-meson decays, the efficiency is around \percentageNumber{90}, depending on the background level.

%Comparing with the momentum spectrum shown in \cref{fig:pt_distribution},
The efficiency for most of the charged particles expected at typical \BelleII collisions is higher than \percentageNumber{93} for up to two times the expected beam background.
This is an integrated number, which is effectively a convolution of the efficiency as a function of  $p_\mathrm{T}$  and the  $p_\mathrm{T}$  spectrum  for \PUpsilonFourS{}~events.
Tracks with transverse momenta below \SI{100}{\mega\electronvolt\per c} impose complex problems to the track finding due to the small number of hits, high multiple scattering and the high level of background in the innermost layers.
As a result, the efficiency decreases.
The difference between the non-background and the expected beam background is small.

\begin{figure}
  \centering
  \subfloat[\label{fig:performance:svdimpact:pt}]{
    \includegraphics[scale=0.355]{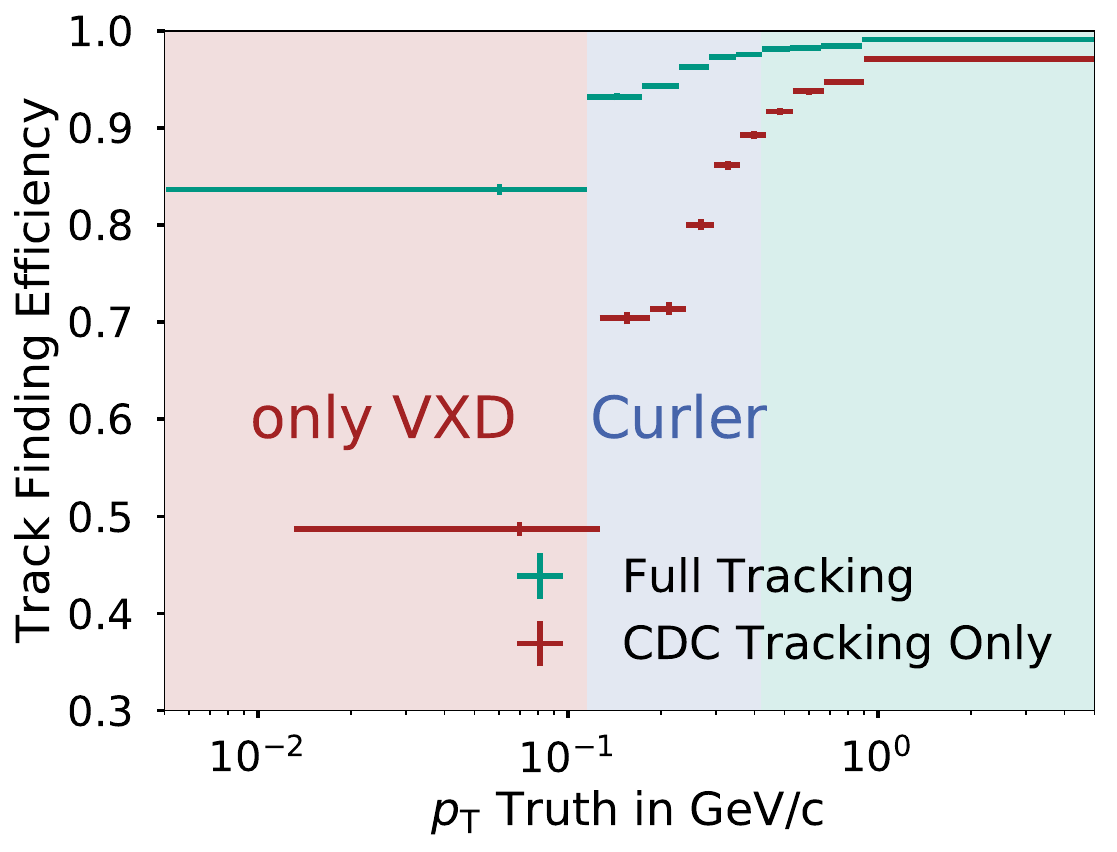}
  }
  \subfloat[\label{fig:performance:svdimpact:theta}]{
    \includegraphics[scale=0.355]{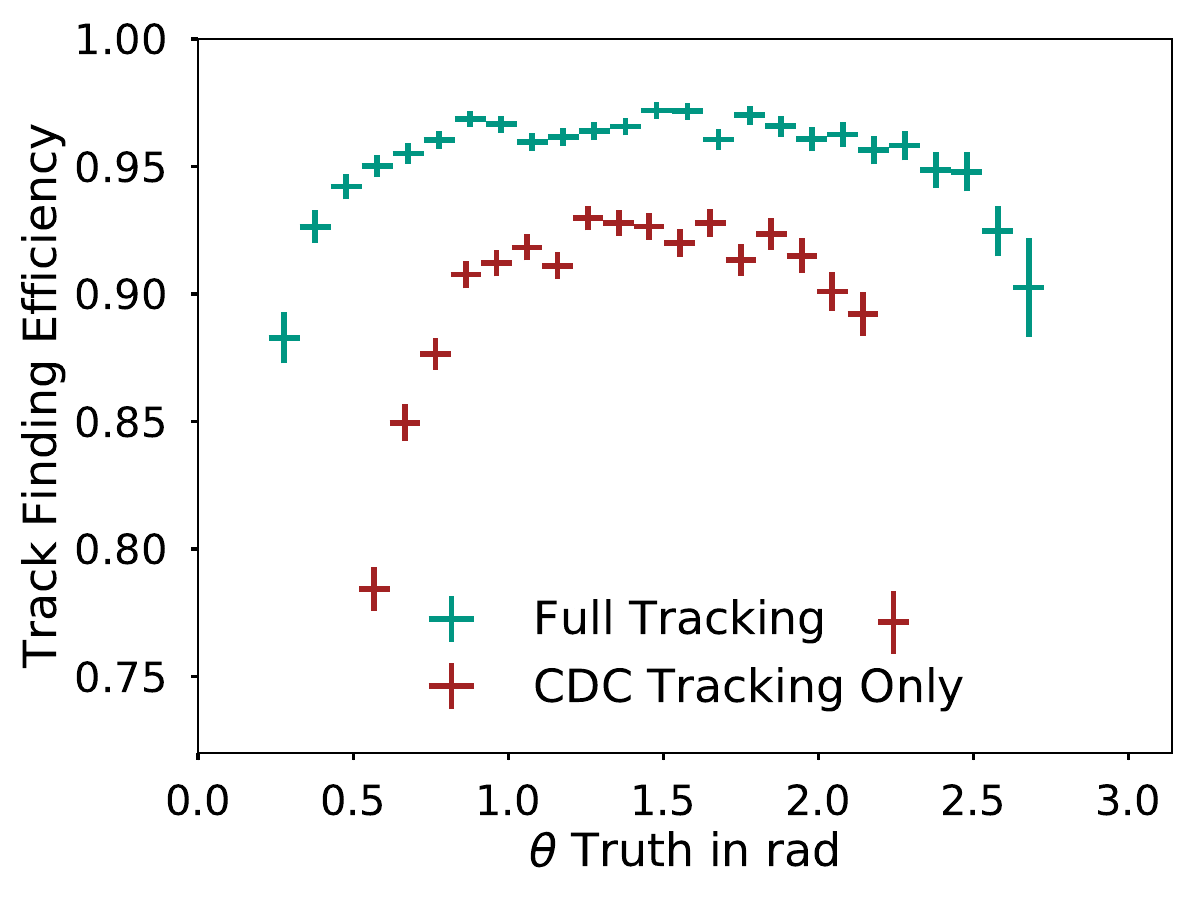}
  }
  \caption{
    Improvement of the track finding efficiency due to the information obtained with the SVD for nominal background.
    The dependence of the track finding efficiency on
    (a) the transverse momentum and
    (b) the polar angle is shown for the full track finding algorithm chain and the CDC standalone algorithm.
  }
  \label{fig:performance:svdimpact}
\end{figure}

The importance of the track finding algorithms which make use of the information of the SVD is illustrated in \cref{fig:performance:svdimpact}.
The plot shows a comparison of the track finding efficiencies for the CDC track finding algorithm and the full chain of tracking algorithms for primary particles and based on a data set with nominal background.
As expected, the impact of the information from the SVD is particular strong in the low momentum region, which can be seen in \cref{fig:performance:svdimpact:pt} where the track finding efficiency dependence on the transverse momentum is shown.
An improvement for curling tracks is also clearly visible.
The better coverage of the acceptance along the polar angle $\theta$ due to the addition of the SVD is well visible in \cref{fig:performance:svdimpact:theta}.

\begin{figure}
  \centering
  \subfloat[
    Track finding efficiency extracted for the most important particle types present in the primary $\PB\APB$-decays at \BelleII dependent on the beam-induced background level.
    The overall finding efficiency is dominated by pions.
    \label{fig:performance:fe:type}
  ]{
    \centering
    \includegraphics[scale=0.32]{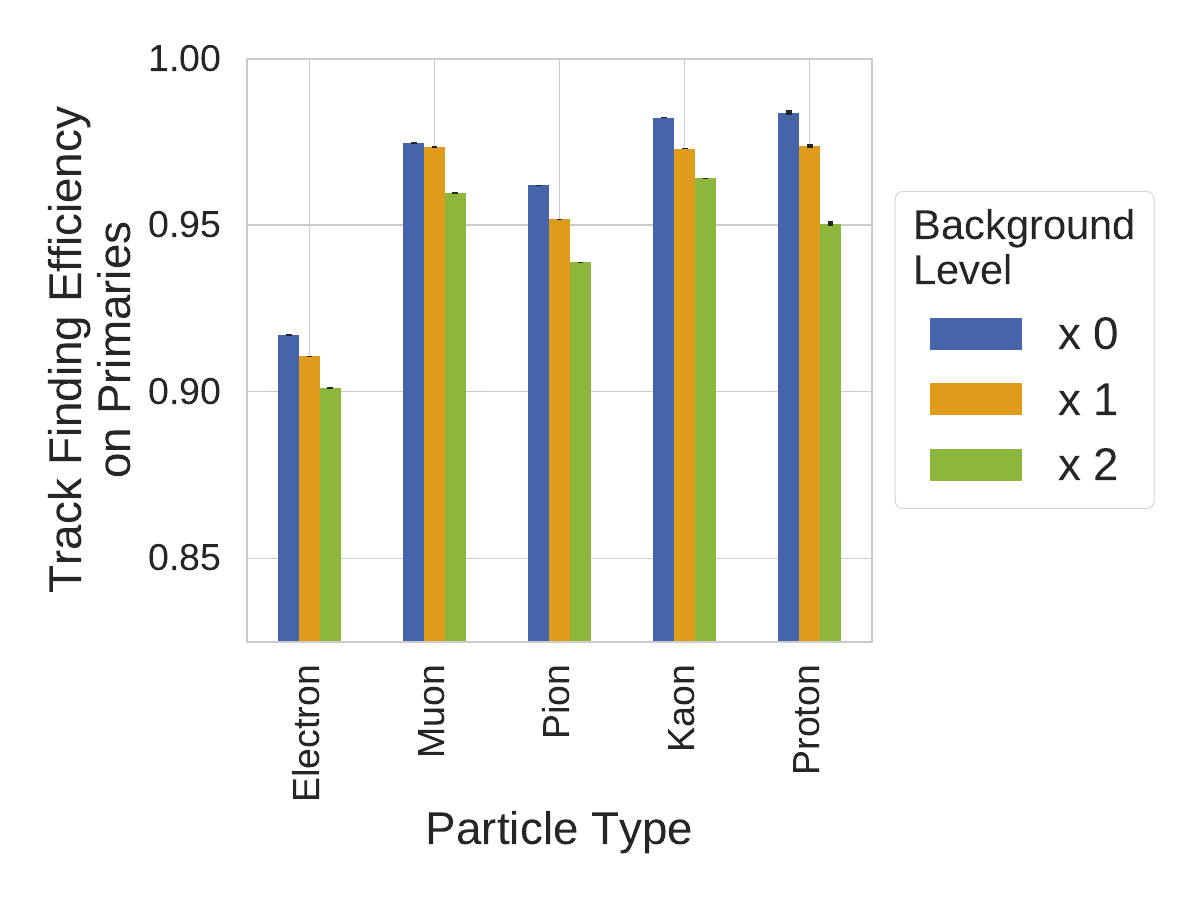}
  }
  \hspace*{1em}
  \subfloat[
    Calculated finding efficiency for those tracks, which are also successfully fitted by the DAF algorithm dependent on the beam-background level.
    The difference with respect to \cref{fig:performance:fe:prim} is negligible.
    \label{fig:performance:fe:fit}
  ]{
    \centering
    \includegraphics[scale=0.32]{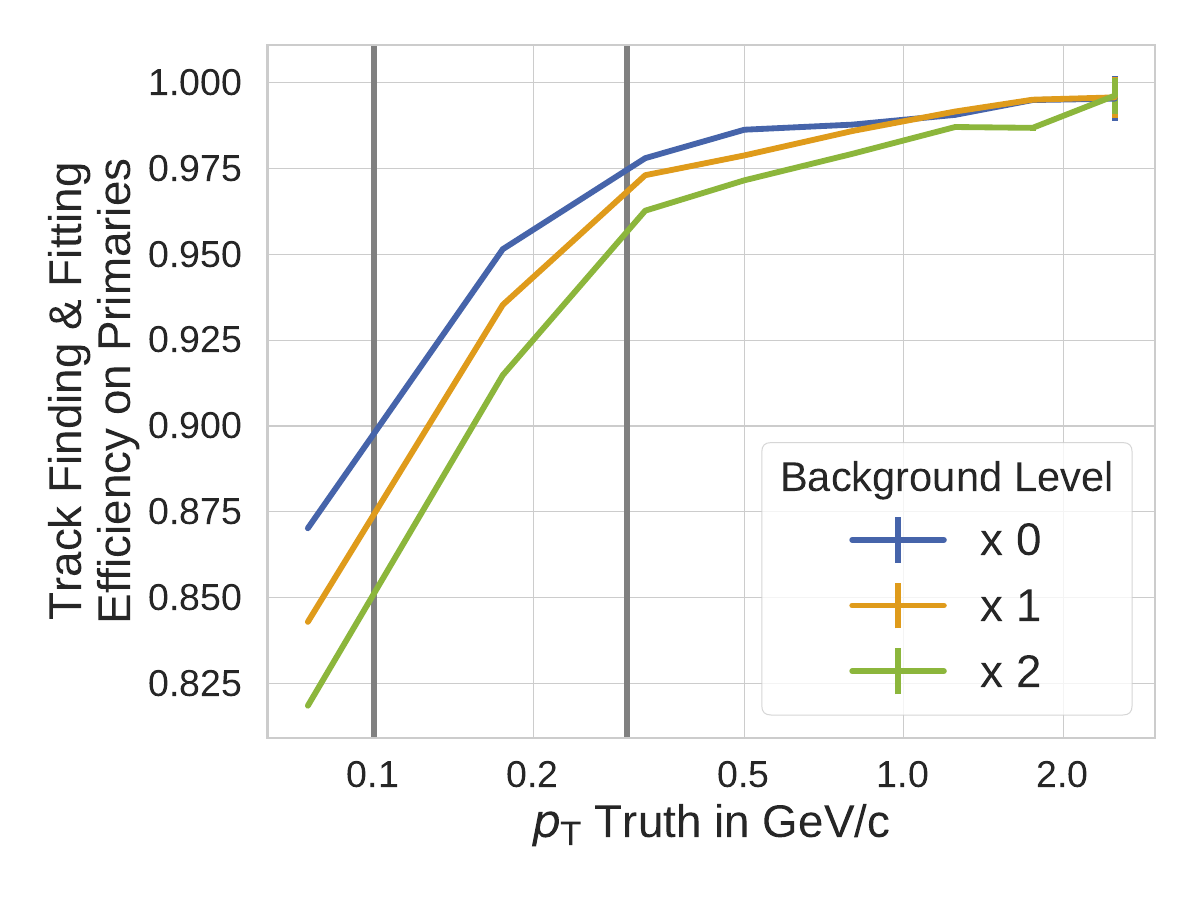}
  }
  \caption{Finding efficiency by particle type and combined finding and fitting efficiency.}
\end{figure}

In \cref{fig:performance:fe:type} the finding efficiency on primaries is compared for different simulated particle types.
Due to the different interaction of electrons with the material, their trajectories are more likely to differ from the nominal helical path, making their reconstruction more challenging.
However, the \BelleII algorithms are able to achieve high efficiencies for every shown particle type for up to twice the expected beam background level.

After the final fit with the DAF provided by \genfit{} the tracks are extrapolated to the POCA to the origin to extract their helix parameters.
In the following, only the results for the pion hypothesis are shown, as most of the produced charged final-state particles are pions.

As a first result, \cref{fig:performance:fe:fit} shows the finding efficiency calculated only with those tracks, where the \genfit{} fit converged and the extrapolation succeeded.
As expected, the difference to \cref{fig:performance:fe:prim} is negligible demonstrating that the track finding algorithms deliver high-quality track candidates to the fitting algorithm.

The helix parameters of the tracks at the POCA can then be compared to the MC truth values.
The resolution $r_x$ is calculated as the \percentageNumber{68} coverage of the the residual $x$ between reconstructed and truth value given as
$$r_x = P_{68\,\%}\left( \left\vert x - P_{50\,\%}(x) \right\vert \right) \ ,$$
where $P_{q}$ calculates the $q$-th percentile of a distribution. Note that $P_{50\,\%}(x)$ corresponds to
the median of $x$.
For a Gaussian distribution, the \percentageNumber{68} coverage and the standard deviation agree.
For non-Gaussian distributions the coverage is more robust against outliers.
As only the results calculated with the pion hypothesis are shown, only true pions from \PUpsilonFourS{}~decays are taken into account for this study.

\begin{figure}
  \centering
  \includegraphics[scale=0.40]{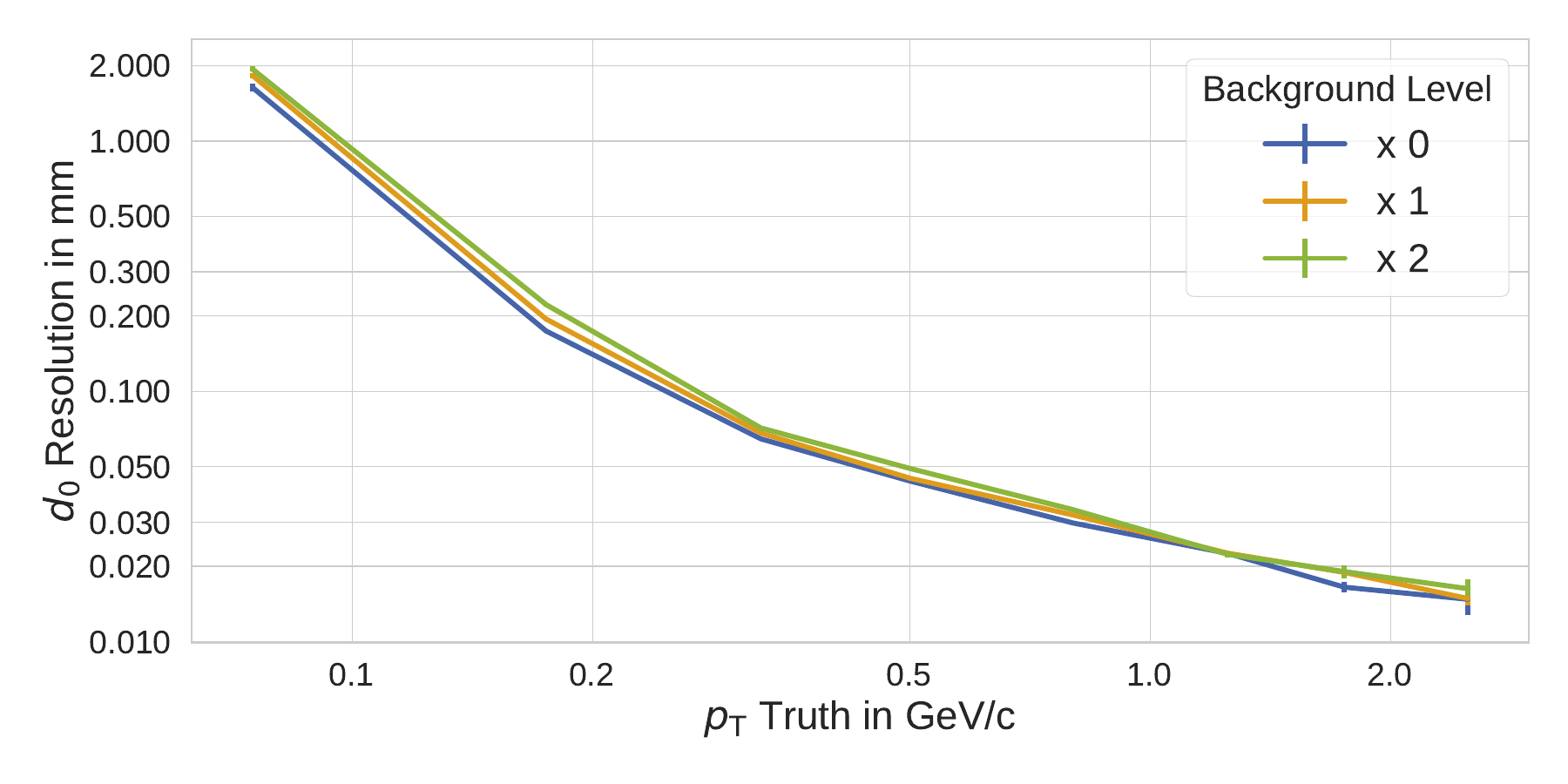}
  \includegraphics[scale=0.40]{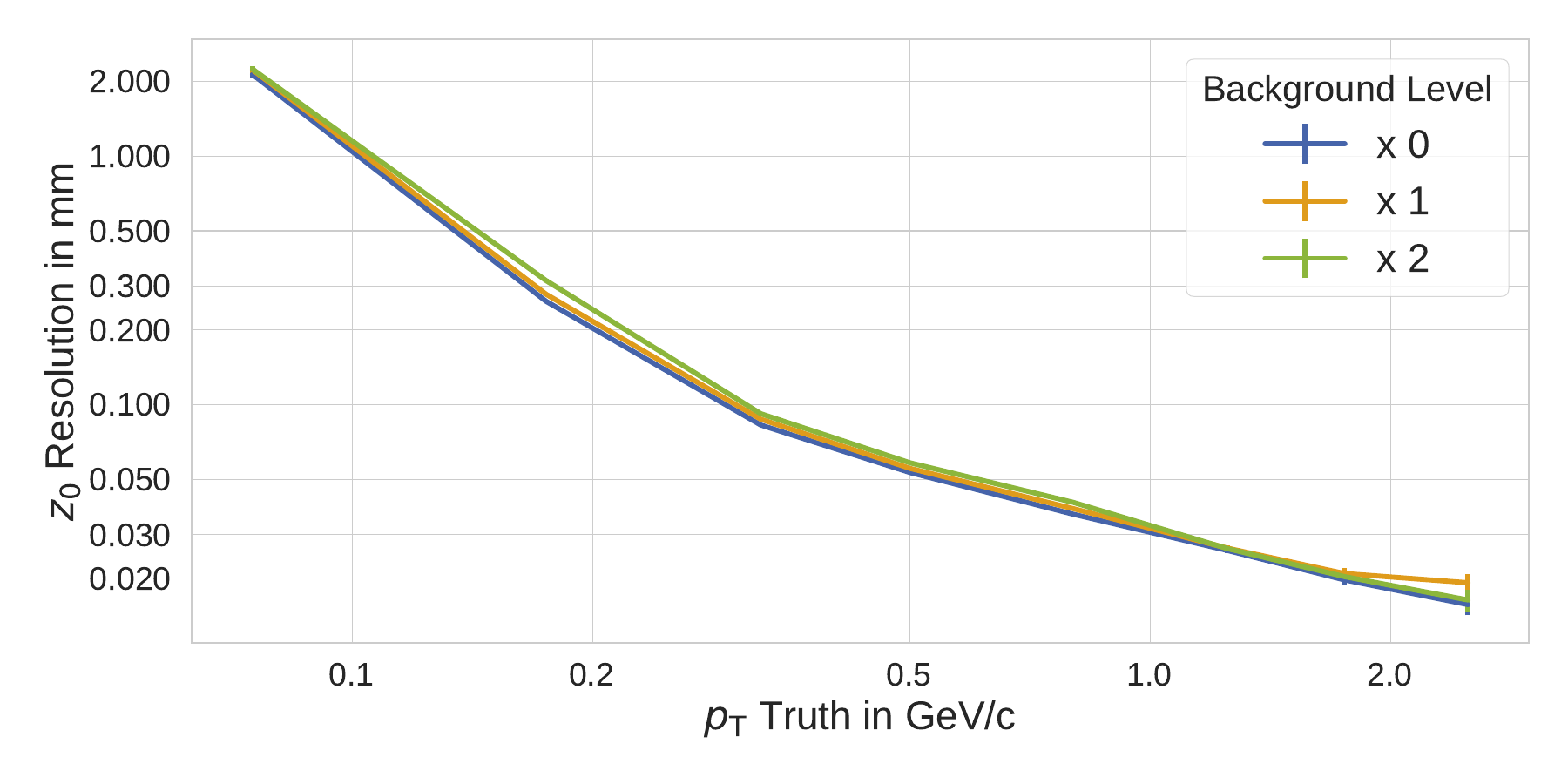}
  \includegraphics[scale=0.40]{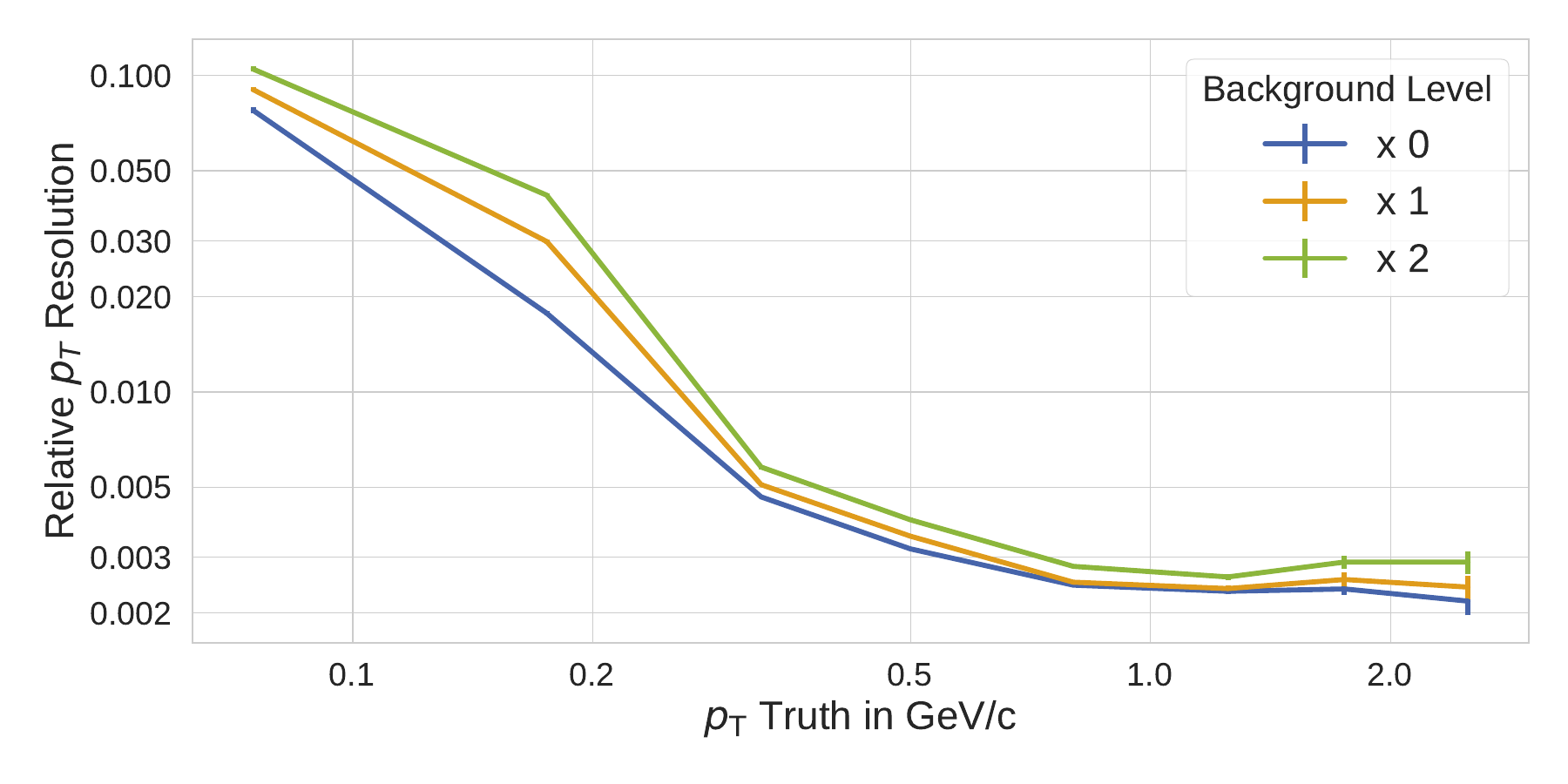}
  \caption{
    Resolutions for typical simulated \PUpsilonFourS{}~events with different levels of beam-induced background.
    As most of the simulated final state particles are pions, only the results of this fit hypothesis are shown.
  }
  \label{fig:performance:resolution}
\end{figure}

In \cref{fig:performance:resolution}, the resolution as a function of the truth transverse momentum is shown for the helix parameters $d_0$ and $z_0$, and for $p_T$ (the transverse momentum).
Both of the spatial parameters, $d_0$ and $z_0$, are mainly influenced by the precise PXD measurements.
Due to the application of the CKF in the PXD and the combination of the VXDTF2 and the CKF for the SVD, a high precision (which is almost independent of the background level) is achieved.
The resolution of the extracted transverse momentum follows expectation: as smaller momenta are more strongly influenced by multiple scattering and a smaller number of measurable hits in the detector,
the resolution decreases with smaller transverse momenta.

\begin{figure}
  \centering
  \subfloat[For $\PBp\PBm$ events.]{
    \includegraphics[scale=0.35]{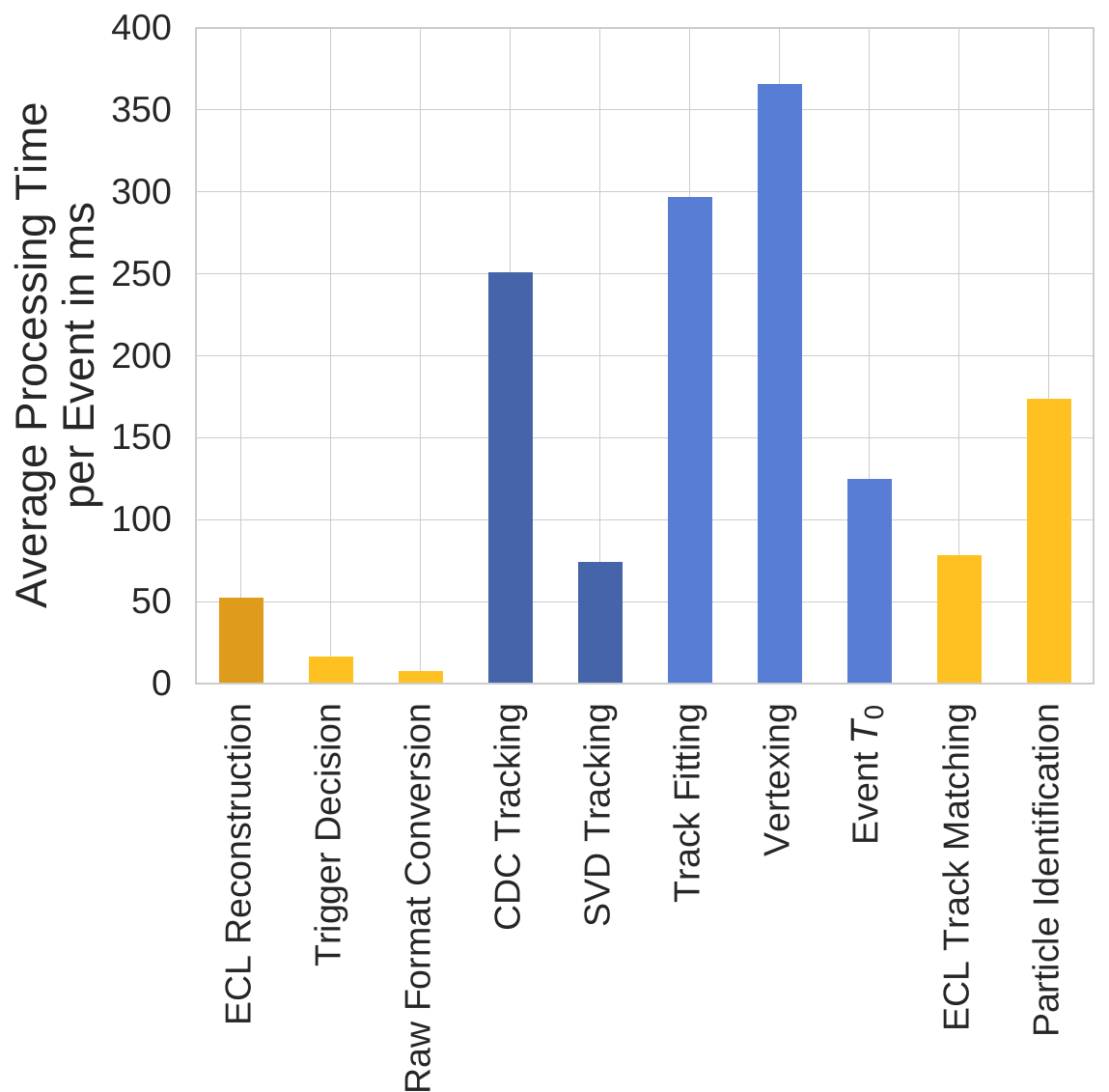}
  }
  \subfloat[For $\Pem\Pep$ events.]{
    \includegraphics[scale=0.35]{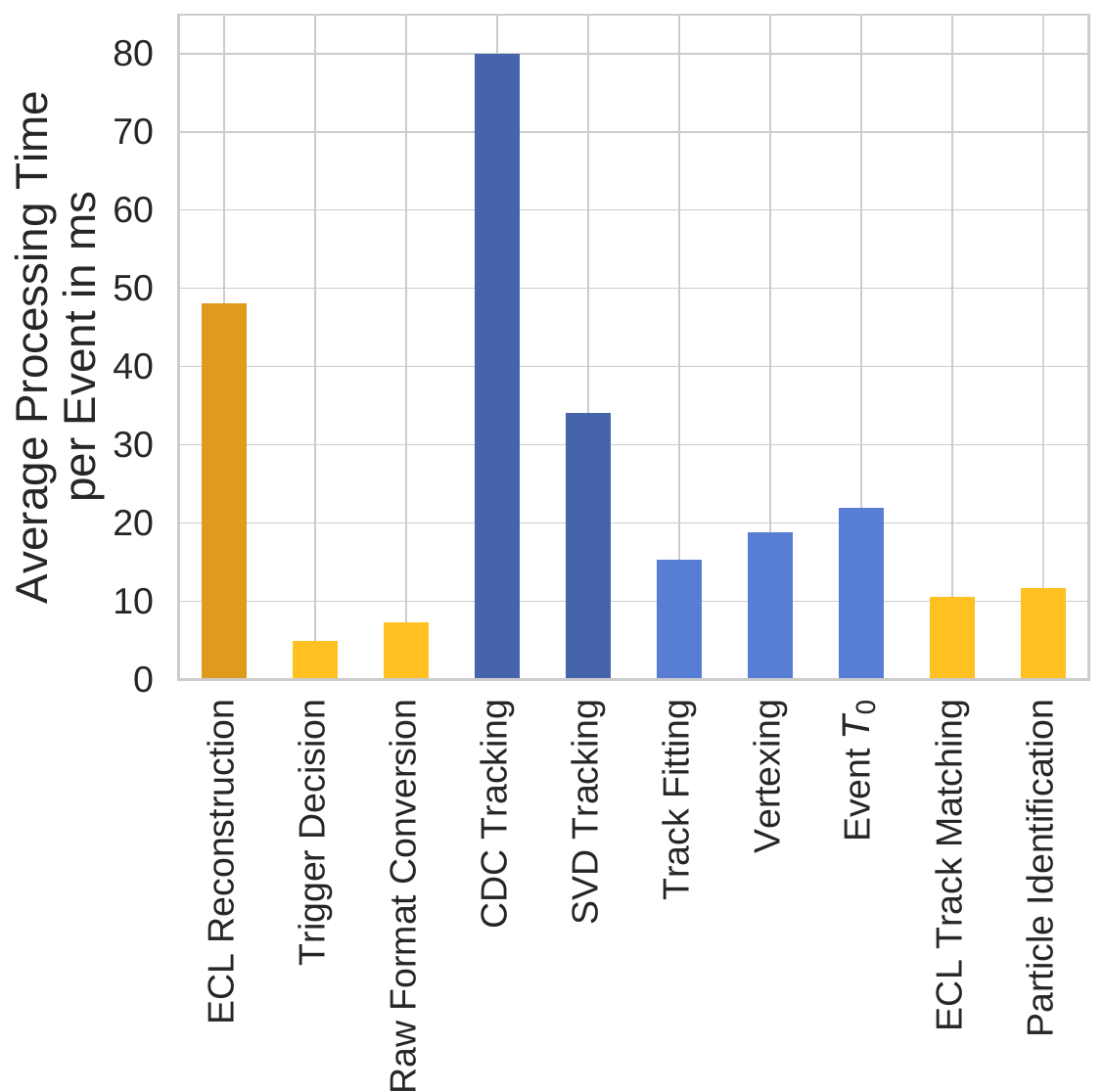}
  }
  \caption{
    Processing time of a standard online reconstruction performed on the anticipated HLT worker nodes in single-processing mode for different event types.
    The components marked in dark (light) blue can be related to track finding (track fitting and other tracking-related tasks).
    The abbreviation \emph{ECL} refers to the electromagnetic calorimeter of the \BelleII{} experiment.
  }
  \label{fig:performance:timing}
\end{figure}

Tracking is one of the most complex tasks in the reconstruction.
It therefore requires a large fraction of the processing time allocated for the online reconstruction on the HLT.
In \cref{fig:performance:timing} the processing time of different components of the online reconstruction performed on one of the HLT worker nodes is shown.
Due to the higher number of tracks in \PUpsilonFourS{}~events, tracking takes longer in this category.
The track fitting, vertexing, and the track-based collision time ($T_0$) extraction are heavily influenced by the handling of the detector geometry in the software.
Different techniques are planned to further optimize the time spent in the geometry navigation.
This is expected to decrease the total processing time significantly which would allow to introduce additional higher level algorithms for the HLT decision.
However, even with the large contribution to the total processing time from tracking, a stable reconstruction on the HLT has been achieved.

% !TeX encoding = UTF-8
% !TeX spellcheck = en_US
% !TeX root = ./TrackingPaper.tex

\section{Summary} \label{sec:summary}
The \BelleII track-reconstruction software consists of multiple independent algorithms to process the measurements of each tracking detector
and integrate all available information into one final set of tracks available for physics analyses.
This allows the use of different algorithms whose properties are especially suited for the three different tracking detectors.
Ultimately, the software provides a set of tracks based on the measurements of all tracking detectors, thereby alleviating the complex task of track combination which would otherwise be forced on physics analyses.

One challenge in this approach is to perform the combination of the tracks reconstructed in each of the tracking detectors without increasing the fake and clone rate.
The best method to achieve this for the \BelleII experiment turned out to be a Combinatorial Kalman Filter to link hits and tracks across detector boundaries.

The upgraded SuperKEKB collider will have much higher beam background radiation and the newly developed tracking algorithms are designed to address this.
Here, multiple methods like early background hit filtering using multivariate methods and a fine-grained candidate selection using the \SectorMap{} concept are used.
The studies with simulated background environments at design luminosity of the accelerator show that tracking performance remains adequate for the expected background rate.

The \BelleII track-reconstruction software has been extensively studied and used for the reconstruction of simulated events.
In the years 2018 and 2019, the software was also employed during the first data taking of collision data and performed well for the commissioning of the \BelleII detector,
studies of the background rates, and first physics results.

\section{Acknowledgments}
We would like to acknowledge the contributions of
Karol Adamczyk,
Bozek Andrzej,
Kirill Chilikin,
David Dossett,
Torben Ferber,
Stefan Ferstl,
Renu Garg,
Hadrien Grasland,
Yinghui Guan,
Yoshihito Iwasaki,
Peter Kodys,
Ilya Komarov,
Jo-Frederik Krohn,
Miriam K\"unzel,
Stefano Lacaprara,
Klemens Lautenbach,
Frank Meier,
Moritz Nadler,
Benjamin Oberhof,
Hitoshi Ozaki,
Johannes Rauch,
Jonas Roetter,
Michael Schnell,
Alexei Sibidanov,
Marko Staric,
Jacek Stypula,
Maiko Takahashi,
Umberto Tamponi,
Viktor Trusov,
Makoto Uchida,
Jonas Wagner,
Ian~James Watson,
Jarek Wiechczynski
and
Michael Ziegler
to the development of the \BelleII tracking software.

\bibliography{TrackingPaper}

\begin{thebibliography}{10}
\expandafter\ifx\csname url\endcsname\relax
  \def\url#1{\texttt{#1}}\fi
\expandafter\ifx\csname urlprefix\endcsname\relax\def\urlprefix{URL }\fi
\expandafter\ifx\csname href\endcsname\relax
  \def\href#1#2{#2} \def\path#1{#1}\fi

\bibitem{Ohnishi2013}
Y.~Ohnishi, T.~Abe, T.~Adachi, et~al.,
  \href{https://academic.oup.com/ptep/article-lookup/doi/10.1093/ptep/pts083}{{Accelerator
  design at SuperKEKB}}, Progress of Theoretical and Experimental Physics
  2013~(3) (2013) 3A011--0.
\newblock \href {http://dx.doi.org/10.1093/ptep/pts083}
  {\path{doi:10.1093/ptep/pts083}}.
\newline\urlprefix\url{https://academic.oup.com/ptep/article-lookup/doi/10.1093/ptep/pts083}

\bibitem{Brodzicka2012}
J.~Brodzicka, T.~Browder, P.~Chang, et~al.,
  \href{https://academic.oup.com/ptep/article-lookup/doi/10.1093/ptep/pts072}{{Physics
  achievements from the Belle experiment}}, Progress of Theoretical and
  Experimental Physics 2012~(1) (2012) 4D001--0.
\newblock \href {http://arxiv.org/abs/1212.5342} {\path{arXiv:1212.5342}},
  \href {http://dx.doi.org/10.1093/ptep/pts072}
  {\path{doi:10.1093/ptep/pts072}}.
\newline\urlprefix\url{https://academic.oup.com/ptep/article-lookup/doi/10.1093/ptep/pts072}

\bibitem{Abe2010}
T.~Abe, I.~Adachi, K.~Adamczyk, et~al.,
  \href{http://arxiv.org/abs/1011.0352}{{Belle II Technical Design
  Report}}\href {http://arxiv.org/abs/1011.0352} {\path{arXiv:1011.0352}}.
\newline\urlprefix\url{http://arxiv.org/abs/1011.0352}

\bibitem{Kuhr2019}
T.~Kuhr, C.~Pulvermacher, M.~Ritter, T.~Hauth, N.~Braun,
  \href{http://link.springer.com/10.1007/s41781-018-0017-9}{{The Belle II Core
  Software}}, Computing and Software for Big Science 3~(1) (2019) 1.
\newblock \href {http://arxiv.org/abs/1809.04299} {\path{arXiv:1809.04299}},
  \href {http://dx.doi.org/10.1007/s41781-018-0017-9}
  {\path{doi:10.1007/s41781-018-0017-9}}.
\newline\urlprefix\url{http://link.springer.com/10.1007/s41781-018-0017-9}

\bibitem{Gardner1970au}
M.~Gardner, {Mathematical Games -- The fantastic combinations of John Conway's
  new solitair game life}, Scientific American 223(4) (1970) 120--123.
\newblock \href {http://dx.doi.org/10.1038/scientificamerican1070-120}
  {\path{doi:10.1038/scientificamerican1070-120}}.

\bibitem{glazov:1993au}
A.~Glazov, I.~Kisel, E.~Konotopskaya, G.~Ososkov, {Filtering tracks in discrete
  detectors using a cellular automaton}, Nucl. Instrum. Meth. A329 (1993)
  262--268.

\bibitem{Abt:2002he}
I.~Abt, I.~Kisel, S.~Masciocchi, D.~Emelyanov, {CATS: A cellular automaton for
  tracking in silicon for the HERA-B vertex detector}, Nucl. Instrum. Meth.
  A489 (2002) 389--405.
\newblock \href {http://dx.doi.org/10.1016/S0168-9002(02)00790-8}
  {\path{doi:10.1016/S0168-9002(02)00790-8}}.

\bibitem{Funke:2014dga}
D.~Funke, T.~Hauth, V.~Innocente, et~al., {Parallel track reconstruction in CMS
  using the cellular automaton approach}, J. Phys. Conf. Ser. 513 (2014)
  052010.
\newblock \href {http://dx.doi.org/10.1088/1742-6596/513/5/052010}
  {\path{doi:10.1088/1742-6596/513/5/052010}}.

\bibitem{Alexopoulos2008}
T.~Alexopoulos, M.~Bachtis, E.~Gazis, G.~Tsipolitis,
  \href{https://www.sciencedirect.com/science/article/pii/S0168900208005780
  https://linkinghub.elsevier.com/retrieve/pii/S0168900208005780}{{Implementation
  of the Legendre Transform for track segment reconstruction in drift tube
  chambers}}, Nuclear Instruments and Methods in Physics Research Section A:
  Accelerators, Spectrometers, Detectors and Associated Equipment 592~(3)
  (2008) 456--462.
\newblock \href {http://dx.doi.org/10.1016/j.nima.2008.04.038}
  {\path{doi:10.1016/j.nima.2008.04.038}}.
\newline\urlprefix\url{https://www.sciencedirect.com/science/article/pii/S0168900208005780
  https://linkinghub.elsevier.com/retrieve/pii/S0168900208005780}

\bibitem{Mankel1999}
R.~Mankel, A.~Spiridonov,
  \href{http://linkinghub.elsevier.com/retrieve/pii/S0168900299000133}{{The
  Concurrent Track Evolution algorithm: extension for track finding in the
  inhomogeneous magnetic field of the HERA-B spectrometer}}, Nuclear
  Instruments and Methods in Physics Research Section A: Accelerators,
  Spectrometers, Detectors and Associated Equipment 426~(2-3) (1999) 268--282.
\newblock \href {http://dx.doi.org/10.1016/S0168-9002(99)00013-3}
  {\path{doi:10.1016/S0168-9002(99)00013-3}}.
\newline\urlprefix\url{http://linkinghub.elsevier.com/retrieve/pii/S0168900299000133}

\bibitem{Mankel2004}
R.~Mankel, \href{http://arxiv.org/abs/physics/0402039}{{Pattern recognition and
  event reconstruction in particle physics experiments}}, Reports on Progress
  in Physics 67~(4) (2004) 553--622.
\newblock \href {http://arxiv.org/abs/0402039} {\path{arXiv:0402039}}, \href
  {http://dx.doi.org/10.1088/0034-4885/67/4/R03}
  {\path{doi:10.1088/0034-4885/67/4/R03}}.
\newline\urlprefix\url{http://arxiv.org/abs/physics/0402039}

\bibitem{CMSCollaboration2014}
{CMS Collaboration}, \href{http://arxiv.org/abs/1405.6569}{{Description and
  performance of track and primary-vertex reconstruction with the CMS
  tracker}}, Journal of Instrumentation 9~(10) (2014) P10009--P10009.
\newblock \href {http://arxiv.org/abs/1405.6569} {\path{arXiv:1405.6569}},
  \href {http://dx.doi.org/10.1088/1748-0221/9/10/P10009}
  {\path{doi:10.1088/1748-0221/9/10/P10009}}.
\newline\urlprefix\url{http://arxiv.org/abs/1405.6569}

\bibitem{Aaboud}
{ATLAS Collaboration}, \href{http://arxiv.org/abs/1704.07983}{{Performance of
  the ATLAS Track Reconstruction Algorithms in Dense Environments in LHC Run
  2}}, The European Physical Journal C 77~(10) (2017) 673.
\newblock \href {http://arxiv.org/abs/1704.07983} {\path{arXiv:1704.07983}},
  \href {http://dx.doi.org/10.1140/epjc/s10052-017-5225-7}
  {\path{doi:10.1140/epjc/s10052-017-5225-7}}.
\newline\urlprefix\url{http://arxiv.org/abs/1704.07983}

\bibitem{keck2016}
T.~Keck, \href{http://arxiv.org/abs/1609.06119}{{FastBDT: A Speed-Optimized
  Multivariate Classification Algorithm for the Belle II Experiment}},
  Computing and Software for Big Science 1~(1) (2017) 2.
\newblock \href {http://arxiv.org/abs/1609.06119} {\path{arXiv:1609.06119}},
  \href {http://dx.doi.org/10.1007/s41781-017-0002-8}
  {\path{doi:10.1007/s41781-017-0002-8}}.
\newline\urlprefix\url{http://arxiv.org/abs/1609.06119}

\bibitem{Kemmer1987}
J.~Kemmer, G.~Lutz,
  \href{https://ac.els-cdn.com/0168900287905183/1-s2.0-0168900287905183-main.pdf?_tid=9c6da4c9-b91f-464b-ae11-89016a3e2531&acdnat=1524476602_881ecbe6d098bc1af5882f526389be5a}{{New
  Detector Concepts}}, Nuclear Instruments and Methods in Physics Research 253
  (1987) 365--377.
\newline\urlprefix\url{https://ac.els-cdn.com/0168900287905183/1-s2.0-0168900287905183-main.pdf?_tid=9c6da4c9-b91f-464b-ae11-89016a3e2531&acdnat=1524476602_881ecbe6d098bc1af5882f526389be5a}

\bibitem{Lewis:2018ayu}
P.~M. Lewis, et~al., {First Measurements of Beam Backgrounds at SuperKEKB},
  Nucl. Instrum. Meth. A914 (2019) 69--144.
\newblock \href {http://arxiv.org/abs/1802.01366} {\path{arXiv:1802.01366}},
  \href {http://dx.doi.org/10.1016/j.nima.2018.05.071}
  {\path{doi:10.1016/j.nima.2018.05.071}}.

\bibitem{Agostinelli2003}
S.~Agostinelli, J.~Allison, K.~Amako, et~al.,
  \href{https://www.sciencedirect.com/science/article/pii/S0168900203013688
  https://ac.els-cdn.com/S0168900203013688/1-s2.0-S0168900203013688-main.pdf?_tid=2ab2887a-7a2b-4813-9966-25cc44845930&acdnat=1524482129_7273334d05ad6f2c66d093ae3c67b132}{{Geant4—a
  simulation toolkit}}, Nuclear Instruments and Methods in Physics Research
  Section A: Accelerators, Spectrometers, Detectors and Associated Equipment
  506~(3) (2003) 250--303.
\newblock \href {http://dx.doi.org/10.1016/S0168-9002(03)01368-8}
  {\path{doi:10.1016/S0168-9002(03)01368-8}}.
\newline\urlprefix\url{https://www.sciencedirect.com/science/article/pii/S0168900203013688
  https://ac.els-cdn.com/S0168900203013688/1-s2.0-S0168900203013688-main.pdf?_tid=2ab2887a-7a2b-4813-9966-25cc44845930&acdnat=1524482129_7273334d05ad6f2c66d093ae3c67b132}

\bibitem{Bilka2019}
T.~Bilka, N.~Braun, T.~Hauth, et~al.,
  \href{http://arxiv.org/abs/1902.04405}{{Implementation of GENFIT2 as an
  experiment independent track-fitting framework}}\href
  {http://arxiv.org/abs/1902.04405} {\path{arXiv:1902.04405}}.
\newline\urlprefix\url{http://arxiv.org/abs/1902.04405}

\bibitem{article:quadtree}
R.~Finkel, J.~Bentley, Quad trees: A data structure for retrieval on composite
  keys., Acta Inf. 4 (1974) 1--9.
\newblock \href {http://dx.doi.org/10.1007/BF00288933}
  {\path{doi:10.1007/BF00288933}}.

\bibitem{Karimaki1991}
V.~Karim{\"{a}}ki,
  \href{https://linkinghub.elsevier.com/retrieve/pii/016890029190533V}{{Effective
  circle fitting for particle trajectories}}, Nuclear Instruments and Methods
  in Physics Research Section A: Accelerators, Spectrometers, Detectors and
  Associated Equipment 305~(1) (1991) 187--191.
\newblock \href {http://dx.doi.org/10.1016/0168-9002(91)90533-V}
  {\path{doi:10.1016/0168-9002(91)90533-V}}.
\newline\urlprefix\url{https://linkinghub.elsevier.com/retrieve/pii/016890029190533V}

\bibitem{Strandlie2000}
A.~Strandlie, J.~Wroldsen, R.~Fr{\"{u}}hwirth, B.~Lillekjendlie,
  \href{https://linkinghub.elsevier.com/retrieve/pii/S0010465500000862}{{Particle
  tracks fitted on the Riemann sphere}}, Computer Physics Communications
  131~(1-2) (2000) 95--108.
\newblock \href {http://dx.doi.org/10.1016/S0010-4655(00)00086-2}
  {\path{doi:10.1016/S0010-4655(00)00086-2}}.
\newline\urlprefix\url{https://linkinghub.elsevier.com/retrieve/pii/S0010465500000862}

\bibitem{Lettenbichler2016}
J.~Lettenbichler, \href{http://www.ub.tuwien.ac.at}{{Real-time Pattern
  Recognition in the Central Tracking Detector of the Belle II Experiment}},
  Ph.D. thesis, Technische Universit{\"{a}}t Wien (2016).
\newline\urlprefix\url{http://www.ub.tuwien.ac.at}

\bibitem{Wagner2017}
J.~Wagner,
  \href{https://ekp-invenio.physik.uni-karlsruhe.de/record/48930/files/EKP-2017-00057.pdf}{{Track
  Finding with the Silicon Strip Detector of the Belle II Experiment}},
  {Master's thesis}, Karlsruhe Institute of Technology (2017).
\newline\urlprefix\url{https://ekp-invenio.physik.uni-karlsruhe.de/record/48930/files/EKP-2017-00057.pdf}

\bibitem{Racs2018}
S.~Racs, \href{https://ekp-invenio.physik.uni-karlsruhe.de/record/49052}{{Track
  Quality Estimation and Defect Sensor Studies for the Silicon Vertex Detector
  of the Belle II-Experiment}}, {Master's thesis}, Karlsruhe Institute of
  Technology, Karlsruhe (2018).
\newline\urlprefix\url{https://ekp-invenio.physik.uni-karlsruhe.de/record/49052}

\bibitem{Fruhwirth2013}
R.~Fr{\"{u}}hwirth, R.~Glattauer, J.~Lettenbichler, W.~Mitaroff, M.~Nadler,
  \href{https://linkinghub.elsevier.com/retrieve/pii/S0168900213008607}{{Track
  finding in silicon trackers with a small number of layers}}, Nuclear
  Instruments and Methods in Physics Research Section A: Accelerators,
  Spectrometers, Detectors and Associated Equipment 732 (2013) 95--98.
\newblock \href {http://dx.doi.org/10.1016/j.nima.2013.06.035}
  {\path{doi:10.1016/j.nima.2013.06.035}}.
\newline\urlprefix\url{https://linkinghub.elsevier.com/retrieve/pii/S0168900213008607}

\bibitem{Berger2017}
N.~Berger, A.~Kozlinskiy, M.~Kiehn, A.~Sch{\"{o}}ning,
  \href{https://linkinghub.elsevier.com/retrieve/pii/S016890021631138X}{{A new
  three-dimensional track fit with multiple scattering}}, Nuclear Instruments
  and Methods in Physics Research Section A: Accelerators, Spectrometers,
  Detectors and Associated Equipment 844 (2017) 135--140.
\newblock \href {http://dx.doi.org/10.1016/j.nima.2016.11.012}
  {\path{doi:10.1016/j.nima.2016.11.012}}.
\newline\urlprefix\url{https://linkinghub.elsevier.com/retrieve/pii/S016890021631138X}

\bibitem{Billoir1989}
P.~Billoir, \href{http://linkinghub.elsevier.com/retrieve/pii/001046558990249X
  https://linkinghub.elsevier.com/retrieve/pii/001046558990249X}{{Progressive
  track recognition with a Kalman-like fitting procedure}}, Computer Physics
  Communications 57~(1-3) (1989) 390--394.
\newblock \href {http://dx.doi.org/10.1016/0010-4655(89)90249-X}
  {\path{doi:10.1016/0010-4655(89)90249-X}}.
\newline\urlprefix\url{http://linkinghub.elsevier.com/retrieve/pii/001046558990249X
  https://linkinghub.elsevier.com/retrieve/pii/001046558990249X}

\bibitem{Lund2009}
E.~Lund, L.~Bugge, I.~Gavrilenko, A.~Strandlie,
  \href{http://stacks.iop.org/1748-0221/4/i=04/a=P04001?key=crossref.4db573cdf47b35d9171ab976705899ed}{{Track
  parameter propagation through the application of a new adaptive
  Runge-Kutta-Nystr{\"{o}}m method in the ATLAS experiment}}, Journal of
  Instrumentation 4~(04) (2009) P04001--P04001.
\newblock \href {http://dx.doi.org/10.1088/1748-0221/4/04/P04001}
  {\path{doi:10.1088/1748-0221/4/04/P04001}}.
\newline\urlprefix\url{http://stacks.iop.org/1748-0221/4/i=04/a=P04001?key=crossref.4db573cdf47b35d9171ab976705899ed}

\bibitem{efron1979}
B.~Efron, \href{http://projecteuclid.org/euclid.aos/1176344552}{{Bootstrap
  Methods: Another Look at the Jackknife}}, The Annals of Statistics 7~(1)
  (1979) 1--26.
\newblock \href {http://dx.doi.org/10.1214/aos/1176344552}
  {\path{doi:10.1214/aos/1176344552}}.
\newline\urlprefix\url{http://projecteuclid.org/euclid.aos/1176344552}

\end{thebibliography}
\bibliographystyle{elsarticle-num}

\end{document}